\documentclass[a4paper, amsfonts, amssymb, amsmath, reprint, showkeys, nofootinbib, twoside, superscriptaddress]{revtex4-1}
\usepackage[english]{babel}
\usepackage[utf8]{inputenc}
\usepackage[colorinlistoftodos, color=green!40, prependcaption]{todonotes}
\usepackage{amsthm}
\usepackage{mathtools}
\usepackage{physics}
\usepackage{xcolor}
\usepackage{graphicx}
\usepackage[left=23mm,right=13mm,top=35mm,columnsep=15pt]{geometry} 
\usepackage{adjustbox}
\usepackage{placeins}
\usepackage[T1]{fontenc}
\usepackage{lipsum}
\usepackage{csquotes}
\usepackage{siunitx}
\usepackage{caption}
\usepackage{subcaption}
\usepackage[pdftex, pdftitle={Article}, pdfauthor={Author}, colorlinks=true]{hyperref}
\begin{document}
\title{
Relaxation of experimental parameters in a Quantum-Gravity Induced Entanglement of Masses Protocol using electromagnetic screening
}
\author{Martine Schut}
    \affiliation{Van Swinderen Institute for Particle Physics and Gravity, University of Groningen, 9747AG Groningen, the Netherlands }
    \affiliation{Bernoulli Institute for Mathematics, Computer Science and Artificial Intelligence, University of Groningen, 9747 AG Groningen, the Netherlands \vspace{1mm}}
    \author{Alexey Grinin}
    \affiliation{Department of Physics and Astronomy, Northwestern University, 2145 Sheridan Road, Evanston, IL}
    \author{Andrew Dana}
    \affiliation{Department of Physics and Astronomy, Northwestern University, 2145 Sheridan Road, Evanston, IL}
    \author{Sougato Bose}
    \affiliation{Department of Physics and Astronomy, University College London, London WC1E 6BT, United Kingdom}
    \author{Andrew Geraci}
    \affiliation{Department of Physics and Astronomy, Northwestern University, 2145 Sheridan Road, Evanston, IL}
\author{Anupam Mazumdar}
    \affiliation{Van Swinderen Institute for Particle Physics and Gravity, University of Groningen, 9747AG Groningen, the Netherlands }


\begin{abstract}
To test the quantum nature of gravity in a lab requires witnessing the entanglement between the two test masses (nano-crystals) solely due to the gravitational interaction kept at a distance in a spatial superposition. The protocol is known as the quantum gravity-induced entanglement of masses (QGEM). 
One of the main backgrounds in the QGEM experiment is electromagnetic (EM) induced entanglement and decoherence. The EM interactions can entangle the two neutral masses via dipole-dipole vacuum-induced interactions, such as the Casimir-Polder interaction. To mitigate the EM-induced interactions between the two nano-crystals, we enclose the two interferometers in a Faraday cage and separate them by a conducting plate. However, any imperfection on the surface of a nano-crystal, such as a permanent dipole moment will also create an EM background interacting with the conducting plate in the experimental box. These interactions will further generate EM-induced dephasing which we wish to mitigate. In this paper, we will consider a parallel configuration of the QGEM experiment, where we will estimate the EM-induced dephasing rate, run-by-run systematic errors which will induce dephasing, and also provide constraints on the size of the superposition in a model-independent way of creating the spatial superposition.
\end{abstract}


\maketitle

\section{Introduction}\label{sec:intro}
Quantum spatial superposition and entanglement~\cite{Horodecki:2009zz} are the two key tools to test the quantum nature of gravity in a laboratory~\cite{Bose:2017nin}, see also~\cite{Marletto:2017kzi}~\footnote{The results of \cite{Bose:2017nin} were first reported in a conference talk \cite{ICTS}.}. Both tools are inherently quantum in nature and there are no classical counterparts to them. In~\cite{Bose:2017nin}, it was pointed out that the two masses of order $m\sim 10^{-14}\,{\rm kg}$ if kept in a spatial superposition of order $100\,{\rm \mu m}$, separated at a distance of $450\,{\rm \mu m}$ for a time $\tau\sim 1-2\,$s will entangle them gravitationally, sufficiently to be detectable via entanglement witness~\cite{Horodecki:2009zz}, if gravity is inherently a quantum entity where the gravitational interactions are treated locally and the pillars of relativistic quantum field theory is maintained~\cite{Marshman:2019sne,Bose:2022uxe,Vinckers:2023grv,Elahi:2023ozf,Carney_2019,Belenchia:2018szb,Danielson:2021egj,Christodoulou:2022mkf}~\footnote{One can introduce non-local gravitational interaction and can compute entanglement witness, see \cite{Marshman:2019sne,Vinckers:2023grv}, but the non-local interaction is introduced at the level of Lagrangian in a very specific non-local, infinite derivative theories of gravity~\cite{Biswas:2011ar,Biswas:2005qr}, such non-locality can be perceived arising from string theory, see e.g.,~\cite{Abel:2019zou} or string field theory~\cite{Witten:1985cc}.}.The protocol introduced in \cite{Bose:2017nin} is known as quantum gravity-induced entanglement of masses (QGEM).

This simple but potent observation made in ~\cite{Bose:2017nin} also supports the principle of local operation and classical communication (LOCC)~\cite{Bennett:1996gf}, which states that if two quantum systems if they were pure, to begin with then classical communication would not entangle them at all. 
One would require a quantum communication/mediator to entangle the two quantum systems~\cite{Bose:2017nin,Marshman:2019sne,Bose:2022uxe}.
We refer to Refs.~\cite{Bennett:1996gf,Bose:2022uxe,Marshman:2019sne,Carney_2019,Christodoulou:2022mkf,Danielson:2021egj} for an extensive discussion of the argument that only quantum interactions can generate entanglement in an initially unentangled state.  
In brief, these references show that interaction via a virtual quantum state is needed for the generation of entanglement in the case of gravity. Virtual states are one of the most bonafide quantum states one can get -- a superposition of all the energy eigenstates. A classical interaction will give only the on-shell states (which are not quantum superposed, they are in a specific state) and can therefore not generate entanglement, see Ref. ~\cite{Bose:2017nin,Marshman:2019sne,Bose:2022uxe}. 

By classical communication, we mean that there is a classical probability associated with the local operations, e.g. Unitary operations, but no Hilbert state associated to classical operators. Therefore, irrespective of formal aspects of quantum gravity or its ultraviolet challenges, i.e. however way we may quantize gravity if gravity is a quantum entity it would entangle the two spatially superposed masses even at the lowest order in the potential. In fact, Newtonian potential has no $\hbar$, and despite this witnessing entanglement between the two superposed masses holds the key to unveiling the quantum nature of gravitational interaction. 

This has been shown via covariant quantization of perturbative quantum gravity in Ref.~\cite{Marshman:2019sne,Bose:2022uxe,Carney_2019}, path integral approach~\cite{Christodoulou:2022mkf}, and by axiomatic approach to perturbative quantum gravity, see Ref.~\cite{Danielson:2021egj}. The only assumptions in all these papers are locality, causality, and relativistic quantum field theory of gravity is applicable around a well-defined Minkowski's background.

This observation is similar in spirit to Bell's test of quantum nonlocality that even if $\hbar \rightarrow 0$ the quantum correlation does not vanish~\cite{Hensen:2015ccp}, first observed in the case of large spins violating Bell's inequality, see Refs.~\cite{PhysRevA.46.4413,GISIN199215}. The QGEM protocol is precisely based on witnessing this correlation via spin entanglement. The idea is to embed spins into the quantum system, such as nitrogen-vacancy (NV) spin in the diamond-type crystal~\cite{Bose:2017nin} and use the spin to create a macroscopic quantum superposition such as in the Stern-Gerlach (SG) apparatus~\cite{PhysRevLett.123.083601,Margalit:2020qcy,Marshman:2021wyk,PhysRevLett.125.023602,Zhou:2022frl,Zhou:2022jug,Zhou:2022epb,Marshman:2023nkh} and close the one-loop interferometer to measure the interference between the two paths, and hence build spin-spin correlations, e.g, entanglement witness,  between the two adjacent interferometers.

The creation of spatial superpositions is experimentally challenging. The system needs to be prepared initially in a pure state~\cite{Whittle:2021mtt,delic2019motional}, and the dephasing factors need to be controlled (as discussed in this manuscript)~\footnote{
Self-gravity effects suggested by Penrose and Diosi~\cite{Penrose:1996cv,DIOSI1987377,Diosi:1989} and investigated in a Stern-Gerlach setup in Ref.~\cite{Sahoo:2022yek} may additionally decohere the superposition. However, Penrose's collapse of the wavefunction demands a violation of the superposition principle in quantum mechanics due to the potential singularity in the theory of general relativity. However, there are theories of gravity motivated from string theory where the gravitational interaction becomes weak at short distances and small time scales can avoid Penrose's conjecture~\cite{Biswas:2005qr,Biswas:2011ar}. In such a class of theories the wavefunction does not collapse even in Diosi's model~\cite{Buoninfante:2017rbw}. Further note that the gravitational binding energy in our case without assuming any modification to the Einstein gravity is given by $U=3Gm^2/5r$. For the experimental parameters used in this manuscript, this is extremely small, e.g. $U\sim10^{-33}\,\si{\joule}$. Thus, $U\ll1$, which is why we can trust the use of an effective quantum field theory to derive the quantum gravitational interaction, see~\cite{Donoghue:1994dn}. Also, for the mass and magnetic field gradient used in this manuscript this hypothetical self-gravitational interaction is not expected to be relevant within the chosen experimental time (based on~\cite{Sahoo:2022yek})}.

Naturally, witnessing the entanglement will be extremely challenging, there are many sources of noise and one particular noise is indeed induced by the electromagnetic (EM) interactions in the neighborhood of the nanocrystals ~\cite{vandeKamp:2020rqh}, decoherence due to heating of the crystal, blackbody emission/absorption, scattering of the ambient quanta~\cite{Bose:2017nin,Chevalier:2020uvv,Nguyen:2019huk,Schut:2021svd,Tilly:2021qef} motivated from~\cite{Romero_Isart_2011,Chang_2009,Sinha:2022snc,RomeroIsart2011LargeQS}. There are also external jitters due to gas molecules, gravity-induced noise such as gravity gradient noise, and relative acceleration noise~\cite{Grossardt:2020def,Toros:2020krn,Wu:2022rdv}, and dephasing due to heavy massive objects near the experiment, e.g., cryogenics and vacuum pump~\cite{Gunnink:2022ner}.

In this paper, we will focus on the EM-induced noise. In particular, we will discuss the situation as first proposed in Ref.~\cite{vandeKamp:2020rqh} where a conducting plate is placed between the two test masses to shield the test masses from interacting vacuum-induced dipole-dipole interaction, e.g., Casimir-Polder potential~\cite{Casimir:1948dh,Casimir:1947kzi}. However, we consider a different geometrical configuration of test masses, namely the test masses are in the `parallel' rather than the `linear' configuration. The reason for selecting such a configuration is to maximize the entanglement phase, and hence the entanglement witness, see~\cite{Nguyen:2019huk,Schut:2021svd,Tilly:2021qef,Rijavec:2020qxd}. Here, we will also consider a wider range of relevant effects, especially focusing on the induced-electric dipole moment, and the dipole moment on the surface of the crystal~\cite{Afek:2021bua,Rider:2016xaq}. Besides these, there are also common mode fluctuations between the two halves of the interferometers. Here also, we will assume that the two halves of the experiment, the two interferometers are separated by a conducting plate to minimize the EM-induced interactions between the nano-crystals, first proposed in~\cite{vandeKamp:2020rqh}. We will analyse 
various sources of dephasing, namely the fluctuation in the paths of the interferometer, run-to-run fluctuations in releasing the nano-crystal's position, fluctuations in the magnetic field, or fluctuations in the conducting plate. All  these fluctuations will manifest in some decoherence, in the sense that they will affect the global phase of the density matrix of the two interferometers, which we wish to optimize for the QGEM experiment.

In sec.~\ref{sec:setup}, we explain why the parallel configuration and the introduction of a conducting plate could be beneficial in terms of witnessing the entanglement.
Then we add a conducting plate to the setup (sec.~\ref{sec:plate}) and discuss the change in the free-fall trajectories of the test masses due to the Casimir-and dipole-interaction between the test masses and the plate (secs.~\ref{subsec:casimir} and~\ref{subsec:dipole}, respectively).
Due to the test masses' trajectories and therefore the accumulated phase being dependent on the initial separation of the test masses to the plate, small fluctuations in the creation of the initial state (such as the distances to the plate, the magnetic gradient, and the dipole orientation) will result in phase fluctuations at the measurement stage. This is discussed in sec.~\ref{sec:phase_fluct}.
Furthermore, an initial setup that is not perfectly symmetric can cause a deflection in the conducting plate and consequently cause dephasing in the superpositions due to the plate. 
An estimation of the coherence time due to dephasing from the deflection of the plate, and thermal fluctuations of the plate, is given in sec.~\ref{sec:dephasing}.
We conclude by showing the relaxed experimental parameters of the new setup needed to witness the entanglement.

\section{Parallel and  linear setups}\label{sec:setup}
We will consider a nano-crystal with a  spin, in fact, our discussion is very generic and can be applied to many dopants. As an example, we may consider  diamond like system with one NV-center, we will assume that the crystal is a sphere and the NV is at the center, for a review on NV-center diamond, see~\cite{Barry:2019sdg,gali2019emph}.
We will also assume that the crystal is charged neutral, we will take up the case when there are surface dipoles separately~\cite{Afek:2021bua,Rider:2016xaq}.
We will initiate the spin in a superposition state, see below, and let the crystal pass through an inhomogeneous magnetic field of the SG setup, the spin superposition allows for the creation of a spatial superposition, see~\cite{PhysRevLett.123.083601,Margalit:2020qcy,Marshman:2021wyk,article,PhysRevLett.125.023602,Zhou:2022frl}. There are many schemes to create the spatial superposition, but here we will consider a simple setup where we will take into account of three steps, (1) acceleration of the crystal, $\tau_a$, (2) intermediate phase when the crystal is not experiencing any SG force, and (3) the last phase of recombining the trajectories of spin-up and down.
\begin{figure}[ht]
    \centering
    \includegraphics[width=\linewidth]{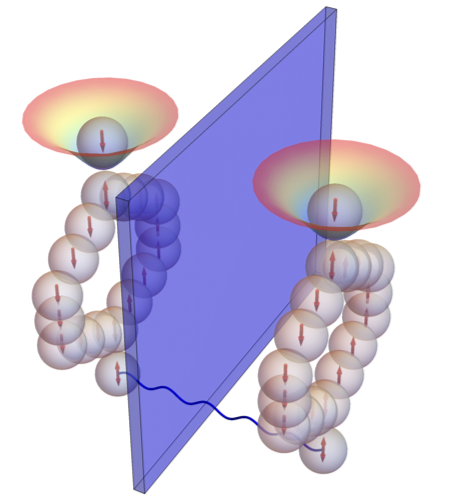}
    \caption{Schematic representation of the anticipated setup and the QGEM protocol. Two nano-crystals with internal spins (red arrows) are initially kept and cooled in two 3D traps separated by a thin membrane. With standard radio-frequency techniques, a spin superposition of each nano-crystal is created $\frac{1}{\sqrt{2}}\left(\ket{0}+\ket{1}\right)$ (double-side arrows). Upon switching off the traps, the spheres fall through a strong B-field gradient region. The two spin states are accelerated in different directions creating spatial superposition states. While falling gravity can entangle the superposition states of the spheres (blue wavy line). Upon passing through an inverted magnetic field the spatial superpositions states are compressed again and the final spin state is detected. This sequence is repeated until enough statistics are gathered to witness gravity-induced entanglement.}
\end{figure}
We distinguish the following two setups that correspond to two different configurations.

\begin{itemize}
 
\item {\it Parallel setup}: The direction in which this superposition is created is such that the two superpositions are parallel, as depicted in figure \ref{fig:setup_par}. It was first considered in~\cite{Tilly:2021qef} and further studied in~\cite{Schut:2021svd} including the effects of decoherence.
\begin{figure}[h]
    \centering
    \begin{tikzpicture}
\draw[black, dashed, thick] (0,-1) -- (0,-0.1);
\draw[black, dashed, thick] (1,-1) -- (1,-0.1);
\draw[black, thick] (0,-1.7) -- (0,-1.9) |- (0,-1.8) -- (1,-1.8) node[pos=0.5, below] {$d$} -| (1,-1.7) -- (1,-1.9);
\draw[black, thick] (-0.6,-1) -- (-0.4,-1) |- (-0.5,-1) -- (-0.5,0) node[pos=0.5, left] {$\Delta x$} -| (-0.6,0) -- (-0.4,0);
\filldraw[black]
(0,-1) circle (2pt) node[align=center, below] {\hspace{2mm}$\ket{0}_1$}
(1,-1) circle (2pt) node[align=center, below] {\hspace{2mm}$\ket{0}_2$};
\draw[black]
(0,0) circle (2pt) node[align=center, above] {\hspace{2mm}$\ket{1}_1$}
(1,0) circle (2pt) node[align=center, above] {\hspace{2mm}$\ket{1}_2$};
\end{tikzpicture}
    \caption{Two test masses of mass $m$, labelled $1$ and $2$, in the parallel configuration. The superposition width is $\Delta x$, and the distance between the $\ket{0}$-states is $d$.}
    \label{fig:setup_par}
\end{figure}
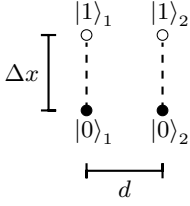
\begin{figure}[h]
\centering
\begin{tikzpicture}
\draw[black, dashed, thick] (-3,0) -- (-2.1,0);
\draw[black, dashed, thick] (-0.5,0) -- (0.4,0);
\draw[black, thick] (-3,-1.3) -- (-3,-1.5) |- (-3,-1.4) -- (-0.5,-1.4) node[pos=0.5, below] {$d$} -| (-0.5,-1.3) -- (-0.5,-1.5);
\draw[black, thick] (-3,-0.7) -- (-3,-0.9) |- (-3,-0.8) -- (-2,-0.8) node[pos=0.5, below] {$\Delta x$} -| (-2,-0.7) -- (-2,-0.9);
\filldraw[black] 
(-3,0) circle (2pt) node[align=center, below] {\hspace{2mm}$\ket{0}_1$}
(-0.5,0) circle (2pt) node[align=center, below] {\hspace{2mm}$\ket{0}_2$};
\draw[black]
(-2,0) circle (2pt) node[align=center, below] {\hspace{2mm}$\ket{1}_1$}
(0.5,0) circle (2pt) node[align=center, below] {\hspace{2mm}$\ket{1}_2$};
\end{tikzpicture}
\caption{Two test masses of mass $m$, labelled $1$ and $2$, in the linear configuration. The superposition width is $\Delta x$, and the distance between the $\ket{0}$-states is $d = d_\text{min} + \Delta x$.}
\label{fig:setup_lin}
\end{figure}

\item{\it linear setup}: The superpositions are kept adjacent to each other, see figure~\ref{fig:setup_lin}. 
In this paper we will focus on the parallel setup, the aim of this section is to show that the parallel results in a larger effective entanglement phase within the first $\sim5\,\si{\s}$ of the experiment.
\end{itemize}

We briefly review the QGEM protocol for  the parallel setup first, a similar analysis on the linear setup has been done earlier, see~\cite{Bose:2017nin,vandeKamp:2020rqh}. We will assume that the initial state of the combined system of the two crystals (labeled system $1$ and system $2$)  is that of a pure state, and  are represented by a spatial superposition of $0$ (spin down) and $1$ (spin up):
\begin{equation}\label{eq:psi0}
    \ket{\Psi_0} 
    =
    \frac{1}{2} \bigotimes_{i=1}^{2} \left( \ket{0}_i + \ket{1}_i \right) \, .
\end{equation}
After creating a spatial superposition, holding it for
a time $\tau$ and then recombining the superposition states, the final state is entangled via the quantum nature of gravity, and the final state is given by:
\begin{equation}\label{eq:psit}
    \ket{\Psi(t=\tau)}
    =
    \frac{e^{i \phi}}{2} \big( \ket{0 \, 0} + e^{i \Delta\phi} \ket{0 \, 1} + e^{i \Delta \phi} \ket{1 \, 0} + \ket{1 \, 1} \big) \, ,
\end{equation}
with~\footnote{Actually, since $\Delta x$ is time-dependent this should be integrated over time. However, for the purpose of this section, we can consider only the entanglement phase generated in a time $\tau$ while $\Delta x$ is constant.}
\begin{align}\label{eq:phase}
    \phi = \frac{G m^2}{d} \frac{\tau}{\hbar}\, , \qq{}
    \Delta \phi = \frac{G m^2}{\sqrt{d^2+(\Delta x)^2}} \frac{\tau}{\hbar} -  \phi \, ,
\end{align}
where $m$ is the mass of the test masses, $d$ and $\Delta x$ are as defined in figure \ref{fig:setup_par}, $G$ is Newton's gravitational constant, $\hbar$ is the reduced Planck's constant.
As long as $2\Delta\phi \neq 2\pi k$ ($k\in\mathbb{Z}$), the state is non-separable (from Pl\"{u}cker's relation~\cite{griffiths2014principles}, or can be seen explicitly in the context of a perturbation theory in quantum mechanics~\cite{Bose:2022uxe}), which means the test masses are entangled, with the maximum entanglement at $\Delta \phi = \pi/2$.
We define therefore the effective entanglement phase as:~\footnote{
In the linear configuration, this effective entanglement phase is:
\begin{equation}
\Phi_\text{eff} = \frac{G m^2}{d+\Delta x} \frac{\tau}{\hbar} + \frac{G m^2}{d-\Delta x} \frac{\tau}{\hbar} -  2 \phi\, . \label{eq:phase_lin}
\end{equation}
}
$$\Phi_\text{eff}=2\Delta \phi \, .$$
Requiring a minimal effective phase, say $\Phi_\text{eff}\sim {\cal O}(1)$ determines the experimental parameters such as the mass $m$ and the distance $d$.
Minimizing the separation would increase the effective phase, but there is a minimal distance $d_\text{min}$ between any two superposition instances required such that the Casimir-Polder-induced entanglement phase is sub-dominant compared to the gravitationally-induced entanglement phase. 
 
\begin{figure}[t]
    \centering
    \includegraphics[width=\linewidth]{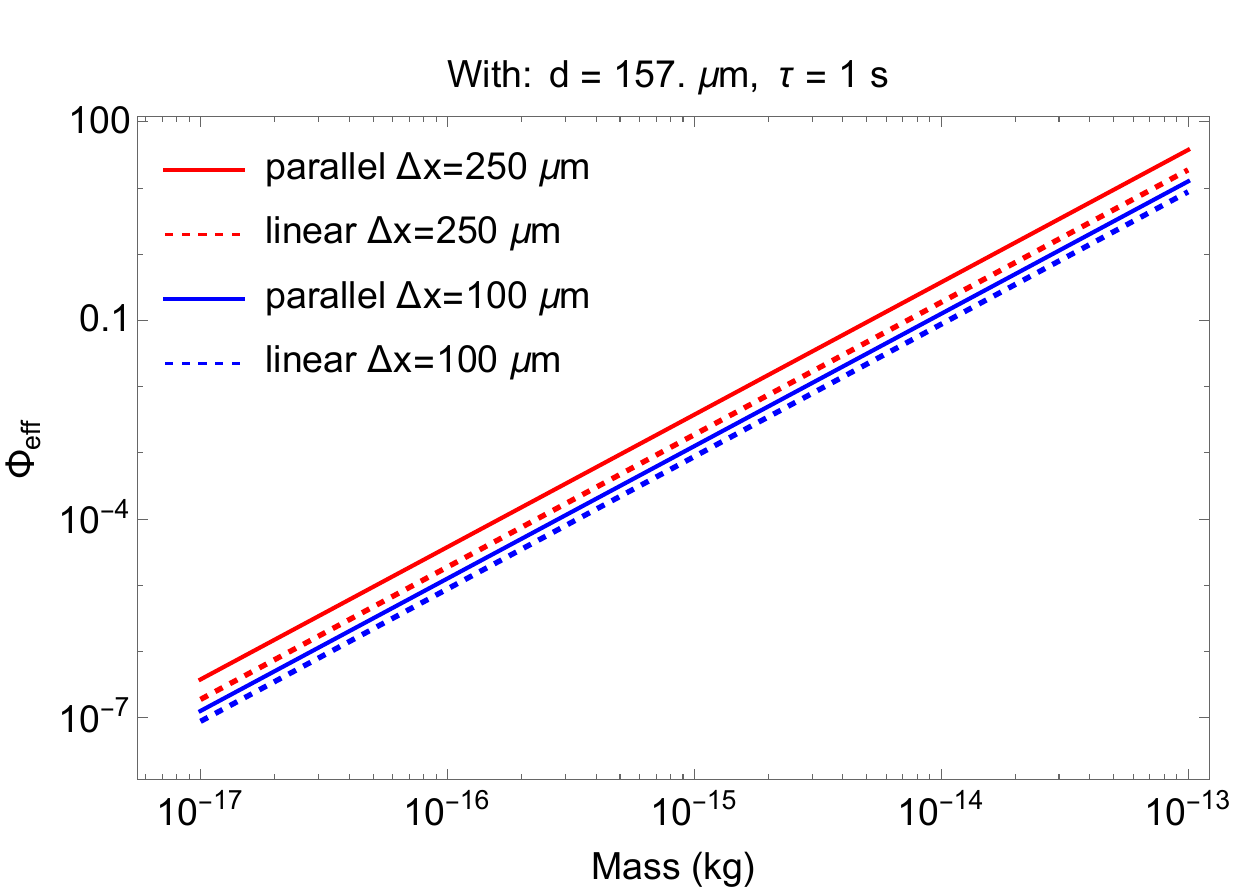}
    \caption{The effective phase of the linear (dashed lines) and parallel (solid lines) configurations as a function of the mass $m$, for $\tau=1 \, \si{\s}$ and $d_\text{min} = 157 \, \si{\micro\metre} $, and for superposition sizes $\Delta x = 100 \, \si{\micro\metre}, 250 \, \si{\micro\metre}$. (colored blue and red, respectively).}
    \label{fig:lin_par_comparison}
\end{figure}

The Casimir-Polder potential between the two neutral dielectric masses is~\cite{Casimir:1948dh,Casimir:1947kzi,vandeKamp:2020rqh,PhysRevLett.125.023602}:
\begin{equation}\label{eq:casimir_spheres}
    V_\text{CP} = - \frac{23\hbar c}{4\pi} \frac{R^6}{d^7} \left(\frac{\varepsilon-1}{\varepsilon+2}\right)^2 \, ,
\end{equation}
with $\varepsilon$ the dielectric constant of the test mass, $r$ the separation of the two states, and $R$ the radius of the test mass. 
Comparing the gravitational - and Casimir-Polder interactions which go as $\sim1/r$ and $\sim1/r^7$, respectively, the Casimir-Polder interaction dominates at short separations. 
From the condition that the gravitational potential is at least one order of magnitude larger than the Casimir-Polder potential, we find a minimal distance (assuming that the test masses are perfect spheres, $\rho=3m/4\pi R^3$)~\citep{vandeKamp:2020rqh}:
\begin{align}
    d \geq \left[ \frac{230}{4\pi} \frac{\hbar c}{G} \left(\frac{3}{4\pi \rho} \frac{\varepsilon-1}{\varepsilon+2} \right)^2\right]^{1/6} \equiv d_\text{min} \approx 157 \mu\text{m}\, ,
\end{align}
where $\rho=3.5\,\si{\g/\centi\metre\cubed}$ (the density of diamond), $\varepsilon=5.1$ (the dielectric constant of diamond), and $c$ the speed of light.
For this minimal distance, we compare the effective entanglement phase for the parallel setup (eq.~\eqref{eq:phase}), with the effective phase from the linear setup (eq.~\eqref{eq:phase_lin}) which was discussed in~\citep{Bose:2017nin,vandeKamp:2020rqh}.
The comparison is given in figure \ref{fig:lin_par_comparison}.
The figure shows that for this minimal distance (which is determined by the material properties), the parallel configuration typically generates a larger effective phase within $1\,\si{\s}$ of interaction, independent of mass~\citep{Nguyen:2019huk,Schut:2021svd}.

Besides the generated entanglement, dephasing via the Casimir interaction also plays a role.
For example, if there is a different Casimir interaction for the two superposition instances between the particles and the conducting plate that will be introduced in the next section.
This specific type of dephasing will be discussed in section~\ref{subsec:fluc_tilt}.
The dephasing due to a general effect can be characterized by the dephasing rate $\gamma_d$.
The final wavefunction presented in eq.~\eqref{eq:psit} including dephasing is~\footnote{Assuming some symmetries in the setup, the equation is derived in more detail in section~\ref{subsec:fluc_tilt}.}:
\begin{align}
    \ket{\Psi(t=\tau)}
    =
    \frac{e^{i \tilde{\phi}}}{2} \big( &\ket{0 \, 0} + e^{i \Delta\phi - i \gamma_d \tau} \ket{0 \, 1} \nonumber \\
    &+ e^{i \Delta \phi} \ket{1 \, 0} + e^{-i\gamma_d \tau} \ket{1 \, 1} \big) \, . \label{eq:psit_dephasing}
\end{align}

\section{Conducting plate}\label{sec:plate}
The previous section shows a parallel orientation of the superposition resulting in a larger entanglement signal compared to a linear configuration, within the first few seconds of the experiment. 
However, Ref.~\citep{vandeKamp:2020rqh} suggested the placement of a conducting plate between the two superpositions, which shields the Casimir-Polder interaction and electric field between the two superpositions, allows for a smaller minimal distance (resulting in a higher entanglement within one second of the total experimental time).
This relaxes the allowed experimental parameters needed to achieve $\Phi_\text{eff}\sim {\cal O} (1)$.
Based on figure \ref{fig:lin_par_comparison} we expect that introducing a conducting plate in the parallel configuration will further aid the parameters.

\begin{figure}[t]
    \centering
    \includegraphics[width=\linewidth]{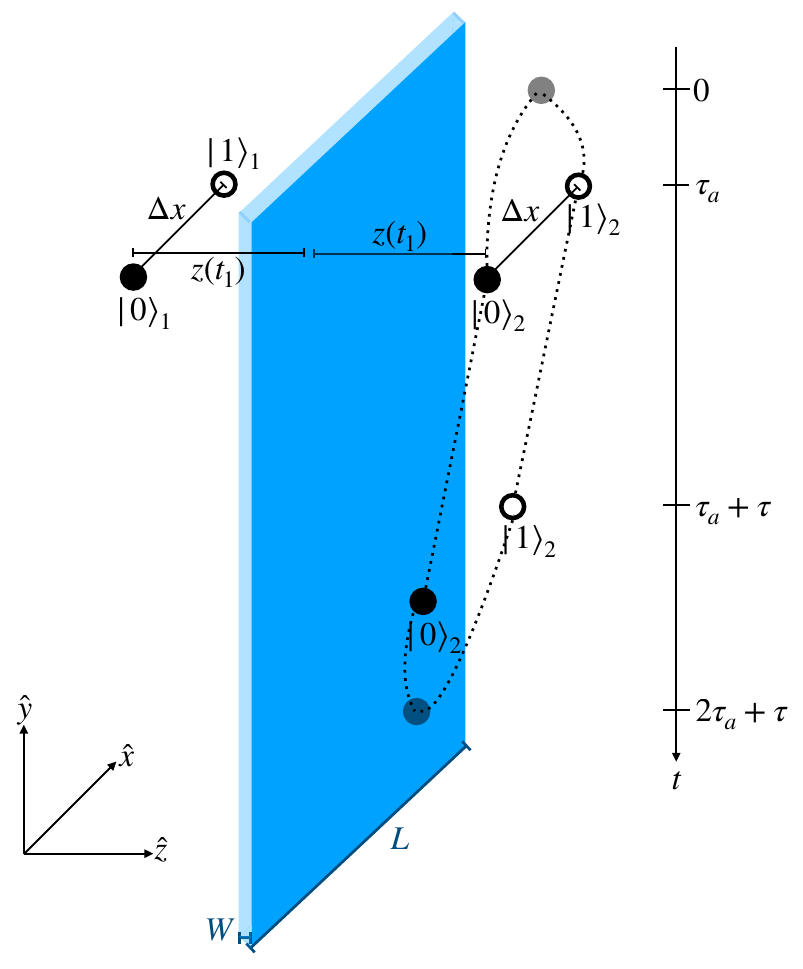}
    \caption{Parallel setup with a conducting plate of size $L$ and width $W$. 
    At $t=0$ there is a spin superposition. 
    By applying a protocol a spatial superposition of size $\Delta x$ is created in $\tau_a$ seconds. After an evolution of $\tau$ seconds, the spatial superpositions are recombined into spin superpositions. 
    The figure shows the maximal superposition width $\Delta x$, and the distance to the plate, $z(t)$ which is time-dependent due to EM interactions with the plate.
    The dotted line illustrates the attraction towards the plate, showing that $z(t)$ decreases over time. We define $z(0)=d$. Note that the figure is not on the scale and that this setup is placed in a Faraday cage of size $L\cross L \cross L$.}
    \label{fig:setup_plate}
\end{figure}

To analyze the parallel configuration, We introduce a perfectly conducting and reflective plate of thickness $W$ at a distance $z$ from both test masses, see figure \ref{fig:setup_plate}.
The plate is assumed to be grounded and is clamped in the $x$-direction with the experimental capsule.
We furthermore assume that the Faraday cage which encloses the experiment is free falling and that the plate and test masses are also in free fall.
The conducting plate will screen the electromagnetic interactions (such as the Casimir-Polder or dipole-dipole interaction) between the two test masses, allowing the minimal distance $d_\text{min}$ to be smaller than in the absence of this plate. However, the conducting plate will interact with the two test masses individually, and hence the resulting force will alter the trajectory of the superposition states by accelerating them toward the plate. 
This will modify the distance to the plate, $z(t)$, over time.
We consider here the Casimir-Polder and the dipole interaction between a dielectric sphere and a conducting plate and show the required initial separation $z(0)=d$ and accumulated effective entanglement phase $\Phi_\text{acc}$ during the experimental time.

Due to the acceleration $a=F/m$ induced by the Casimir/dipole force between the crystal and the conducting plate, within an infinitesimal time period $\delta t$, the change in position of the superposition is modified by a small amount.
Assuming that the acceleration $a$ is constant during an infinitesimal time $\delta t$, the position at $z(t+\delta t)$ is:
\begin{align}
    z(t+\delta t) & = z(t) - \frac{1}{2} a(t) \cdot (\delta t)^2 - v_0(t) \cdot \delta t \, , \label{eq:displacement}
\end{align}
with $v_0(t+\delta t) = v_0(t) + a(t) \cdot \delta t$ the velocity, with initial conditions $v_0(0)=0$, $z(0)=d$.

The effective phase accumulated during an infinitesimal time period $\delta t$ is given by:
\begin{align}
    \Phi_\text{eff}(\delta t) = \frac{2Gm^2\delta t}{\hbar} \bigg[ &\frac{1}{\sqrt{[2z(t+\delta t)+W]^2+(\Delta x)^2}} \nonumber \\ &- \frac{1}{2z(t+\delta t)+W} \bigg] \, . \label{eq:inf_phase}
\end{align}
The total accumulated effective phase can be found by integrating the infinitesimal effective phase over time.
\begin{equation}\label{eq:acc_phase}
    \Phi_\text{acc} = \int_0^\tau \dd{t} \frac{\delta \Phi_\text{eff}}{\delta t} \, .
\end{equation}
Rather than solving this integral analytically, we find the total accumulated phase numerically, by adding the values of eq.~\eqref{eq:inf_phase} over time and updating the position at every time step using eq.~\eqref{eq:displacement}
~\footnote{The python code has been made public on \href{https://github.com/MartineSchut/QGEMShielding}{https://github.com/MartineSchut/QGEMShielding}.}.

So far we have considered only the time period when we can neglect the SG force on the crystal, but we should also include the accumulated phase and the displacement during the creation and recombination of the spatial superpositions.
The superpositions are created by applying a magnetic gradient $\partial_z B$, resulting in the acceleration:
\begin{equation}
    a_m = g \mu_B \partial_z B \, ,
\end{equation}
with the electronic g-factor of the NV-centre $g\sim 2$ and $\mu_B$ the Bohr magneton. The recombination of the spatial superpositions is achieved by simply reversing the direction of the magnetic gradient. 
The direction of the magnetic acceleration $a_m$ is in the $x$-direction and thus perpendicular to the acceleration towards the plate. 
In the parallel configuration, the Casimir and the dipole interactions will not influence the superposition width. The spin-dependent force will be acting along the $x$-direction only.
The displacement due to the SG force, e.g., $\partial B$, is related to the superposition width, and after a time $\tau_a$ will be given by:
\begin{equation}\label{eq:delta_x}
    \Delta x = \frac{g \mu_B \partial_z B}{2 m} \tau_a^2 \, .
\end{equation}
Of course, we are considering a very simple model for creating the superposition, this analysis will give us some idea of how and when the actual experimental setup will be conceived. A more comprehensive protocol for creating a large superposition size within a short time period is discussed in~\cite{Zhou:2022epb,Zhou:2022frl,Zhou:2022jug,Marshman:2023nkh,Marshman:2021wyk}.

The displacement and the infinitesimal phase during the creation/recombination of the spatial superpositions can be found also in Eqs.~\eqref{eq:displacement} and~\eqref{eq:inf_phase}, respectively, but with $\Delta x$ time-dependent and given by eq.~\eqref{eq:delta_x}. 
We first analyze the Casimir-Polder interaction between the conducting plate and the nano-crystal, and then study the surface dipole.


\subsection{Casimir Force}\label{subsec:casimir}

The  Casimir force will act between a diamond-like crystal and the conducting plate. The force between a static plate and a free dielectric sphere is given by~\footnote{
The potential between a dielectric sphere and a perfectly reflective plate was derived by Casimir and Polder in~\cite{Casimir:1947kzi}. In the limit where the separation $z$ is much larger than the wavelength of the electromagnetic field~\cite{Ford:1998ex}:
\begin{equation}\label{eq:casimir_plate_pot}
    V_\text{CP} = - \frac{3 \hbar c \alpha}{8\pi z^4}
\end{equation}
Note that this differs from eq. \eqref{eq:casimir_spheres} which is the potential between two dielectric spheres rather than the potential between one dielectric sphere and a perfectly reflective plate of eq. \eqref{eq:casimir_plate_pot}.
The force due to this interaction was found in~\cite{Ford:1998ex,vandeKamp:2020rqh,PhysRevLett.125.023602}, and can be seen to be $F_\text{CP} = - \partial_z V_\text{CP}$.
We take the complex polarizability of the dielectric sphere, $\alpha=R^3(\varepsilon-1)/(\varepsilon+2)$.
Here we have assumed that the dielectric properties of the test masses are independent of the frequency of the electric field and that its imaginary part is negligible at low temperatures (see experimental findings in~\cite{Floch2011:EPO,Ye2005:die,Ibaraa1997:wide}).
}:~\cite{Ford:1998ex}
\begin{equation}\label{eq:casimir_plate}
    F_c = - \frac{3 \hbar c}{2\pi} \left(\frac{\varepsilon-1}{\varepsilon+2}\right) \frac{3 m}{4 \pi \rho} \frac{1}{z^5} \, ,
\end{equation}
where $z(t)$ is the distance between the plate and the superposition at a given instance of time. We have assumed the crystal to be perfectly spherical, with $\rho$ as the density of the spherical masses ($\frac{4\pi}{3}R^3 \rho = m$).
From eq.~\eqref{eq:displacement}, we can find the position of the dielectric sphere due to the Casimir force-induced displacement via numerical integration. Several methods could be used, we have provided our code on GitHub.

\begin{figure}[b]
    \centering
    \includegraphics[width=\linewidth]{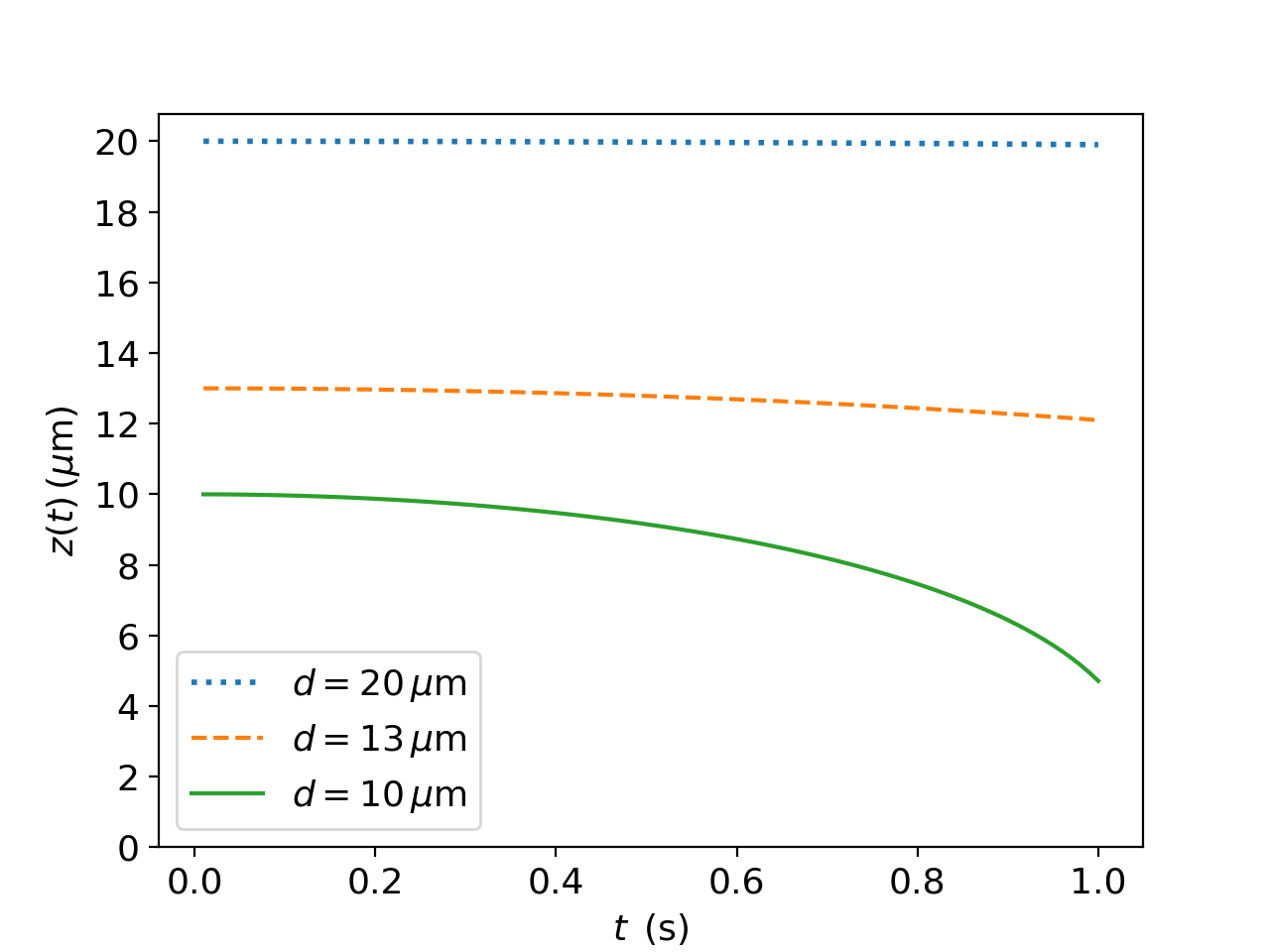}
    \caption{The separation $z(t)$ between one free dielectric sphere and the boundary of the conducting plate changes as a function of time due to Casimir interaction. 
    With the initial separation $z(0)=d$.
    The smallest distance $z(t=1\,s)$ is $20\,\si{\micro\m}$, $12\,\si{\micro\m}$ and $5\,\si{\micro\m}$ for the initial separations of $20\,\si{\micro\m}$, $13\,\si{\micro\m}$ and $10\,\si{\micro\m}$, respectively.}
    \label{fig:trajectory}
\end{figure}

Fig.~\ref{fig:trajectory} shows the trajectory $z(t)$ of a single superposition as a function of time, for different initial values of $z(0) = d$.
Note that this is plot is mass-independent because the mass cancels out in the acceleration.
The closer the initial separation to the plate, the larger the deflection of the trajectory. 
At a separation of around $z(t)\sim15\,\si{\micro\metre}$ the trajectory is approximately constant.


\subsection{Dipole Force}\label{subsec:dipole}
Either due to an external electric field or due to the impurities on the surface of the diamond, the test masses can have some induced or internal dipole moment, respectively~\cite{Afek:2021bua,Rider:2016xaq}.
Using the image method we can find the electric field from the dipole at the surface of the conducting plate. The resulting potential is given by:
\begin{align}\label{eq:dipole_pot}
    V_D = - \vec{p}_1 \cdot \vec{E}_2 = - \Vec{p}_1 \left( \frac{3(\vec{p}_2\cdot \vec{r})\vec{r}}{r^5} - \frac{\vec{p}_2}{r^3} \right) \frac{1}{4\pi\varepsilon_0} \, ,
\end{align}
with $\vec{p}_1$ ($\vec{p}_2$) the dipole moment of the test mass (image test mass) and $\vec{r}$ the radius vector between the two (image) dipoles.
$\vec{E}_2$ is the electric field due to the image dipole $\vec{p}_2$~\cite{griffiths2005introduction,feynman1963feynman}.
The force between the plate and the test mass is then:~\cite{perez2021electric}
\begin{equation}\label{eq:dipole_force}
    F_D = \frac{1}{4\pi\varepsilon_0} \frac{3p^2}{16z^4}[1+\cos^2(\theta)] \, ,
\end{equation}
where $x$ is the separation to the plate, $\theta$ is the angle between the direction of the separation and the dipole moment vector, and $p$ is the dipole moment magnitude.
The dipole moment is given by the sum of the induced and internal dipole moments ($p_e$ and $p_i$, respectively), $p=p_i + p_e$.
The induced dipole due to some external electric field $E_0$ in the dielectric test masses is~\cite{stratton2007electromagnetic,wyld1999math}:
\begin{equation}
    p_e = 4\pi \varepsilon_0 \left( \frac{\varepsilon-1}{\varepsilon + 2} R^3 \right) E_0 \, .
\end{equation}
Considering an external electric field of $E_0\sim 2\times10^{5} \,\si{\metre\kilogram\per\second\cubed\per\ampere}$~\footnote{
As mentioned in section~\ref{sec:intro}, large superpositions can be created using a wire.
The ampacity for copper nanotubes is found to be $J=10^{13}\,\si{\ampere\per\metre\squared}$~\cite{Subramaniam2013:one}.
With a thermal conductivity of $\sigma = 4.6\times10^{7} \,\si{\siemens\per\meter}$ at room temperature~\cite{Subramaniam2013:one}, the electric field found from Ohm's law is $E = J/\sigma \sim 2\times10^{5} \,\si{\metre\kilogram\per\second\cubed\per\ampere}$.\label{footnote:current}
}, 
the externally induced dipole moment is of the order $p_e\sim 6\times10^{-4}\,e\,\si{\centi\metre}$ for $m=10^{-15}\,\si{\kilogram}$. 
The internal dipole moment in the diamond-type crystal is estimated experimentally to be $p_i = 10^{-2} e \,\si{\centi\metre}$ (with $e$ the electric charge) for spheres of $R\sim5\,\si{\micro\metre}$~\cite{Afek:2021bua,Rider:2016xaq}.
A diamond sphere of mass $10^{-16}-10^{-14}\,\si{\kilogram}$ would correspond to $R\approx0.2-0.9\,\si{\micro\metre}$. 
The internal dipole moment of diamond spheres is not exactly known, therefore we take it to be constant at $p_i = 10^{-2} e \,\si{\centi\metre}$, which may be an overestimation.
Since the internal dipole moment for microspheres is two orders of magnitude larger, we continue with just considering a total dipole moment of $p=10^{-2} e \,\si{\centi\metre}$ (at least for the range of masses considered here).

\begin{figure}
    \centering
    \includegraphics[width=\linewidth]{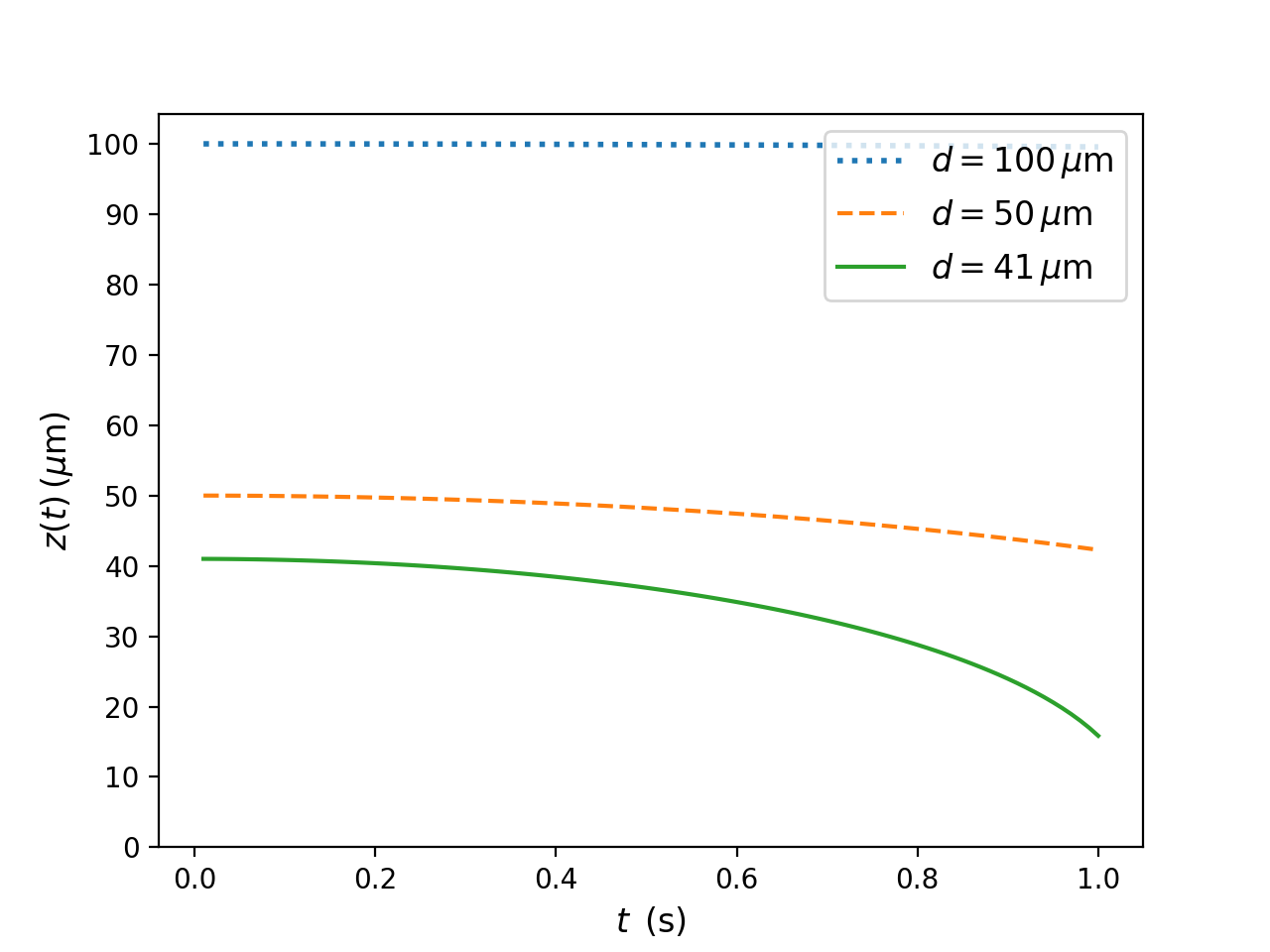}
    \caption{The separation $z(t)$ between one free dielectric sphere and the boundary of the conducting plate changes as a function of time due to dipole interaction. 
    With the initial separation $z(0)=d$. 
    For $m=10^{-14}\,\si{\kg}$. 
    The smallest distance $x(t=1\,s)$ is $100\,\si{\micro\m}$, $42\,\si{\micro\m}$ and $16\,\si{\micro\m}$ for the initial separations of $100\,\si{\micro\m}$, $50\,\si{\micro\m}$ and $41\,\si{\micro\m}$, respectively.}
    \label{fig:trajectory2}
\end{figure}

For an attractive dipole force between the plate and the test mass, given in eq. \eqref{eq:dipole_force}, the position of the test mass (from eq.~\eqref{eq:displacement}) can be found using numerical integration.
The trajectory $z(t)$ of a single superposition instance for different $z(0)=d$ are shown in figure~\ref{fig:trajectory2}.

The displacement due to the dipole interaction is dominant over the displacement resulting from the Casimir interaction~\footnote{
This is the worst-case scenario where the dipoles are aligned such that $\cos(\theta)=1$, and for the estimated $p_i$ mentioned previously.
} as one can see in comparing figures~\ref{fig:trajectory} and~\ref{fig:trajectory2}.
Combining both effects, the initial distance allowed in order for the test mass not to collide with the plate is smaller than the minimal distance required in the absence of the plate. 

\subsection{Accumulated Phase}\label{subsec:acc_phase}
\begin{figure}[t]
    \centering
    \includegraphics[width=\linewidth]{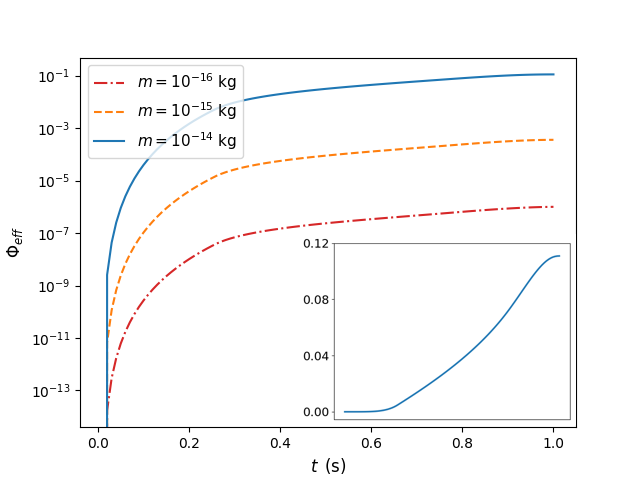}
    \caption{The effective phase accumulated during $1\,\si{\s}$ of experimental time. For masses $m=10^{-16}, 10^{-15}, 10^{-14}\,\si{\kilogram}$ with corresponding $\partial B = 5*10^{3}, 5*10^4, 5*10^5\, \si{\tesla\per\metre}$ such that the superposition width $\Delta x$ is approximately constant at $30\,\si{\micro\metre}$. 
    With $W=1\,\si{\micro\metre}$, $p=10^{-2}e\,\si{\centi\metre}$,  $\tau_a=0.25\,\si{\s}$ and $\tau = 0.5\, \si{\s}$. 
    The initial distance $d$ depends on the mass and is chosen to be $101\,\si{\micro\m}$, $64\,\si{\micro\m}$ and $41\,\si{\micro\m}$for the different masses respectively. 
    The subplot shows the time-dependent phase accumulation during the one second for $m=10^{-14}\,\si{\kilogram}$ without the logarithmic y-axis that is used in the main plot. The closest approach to the plate is approximately $10\,\si{\micro\metre}$.}
    \label{fig:acc_phase}
\end{figure}

The time-dependent accumulated phase is given in eq. \eqref{eq:acc_phase} and plotted in figure~\ref{fig:acc_phase} for the displacement due to the dipole and Casimir interaction.
The dipole is taken to be maximum ($\theta=0$ in eq. \eqref{eq:dipole_force}).
Even after the recombination of the spatial superposition has begun, the phase still grows a lot, which we attribute to the distance between the superpositions decreasing over time, also during this stage of the experiment.
A measurable phase can be achieved by experimentally realizing a superposition width of at least $\sim30\,\si{\micro\metre}$ and closing t within $1\,\si{\s}$ for a mass of $10^{-14}\,\si{\kilogram}$.
A mass that is one order of magnitude smaller, although allowing a smaller separation to the plate due to the smaller attractive force towards the plate, also couples less gravitationally and results in an accumulated effective entanglement phase approximately two orders of magnitude smaller for the same superposition size. 

\section{Entanglement Phase fluctuations from repeated experiments}\label{sec:phase_fluct}
\begin{figure}[t]
    \centering
    \includegraphics[width=0.65\linewidth]{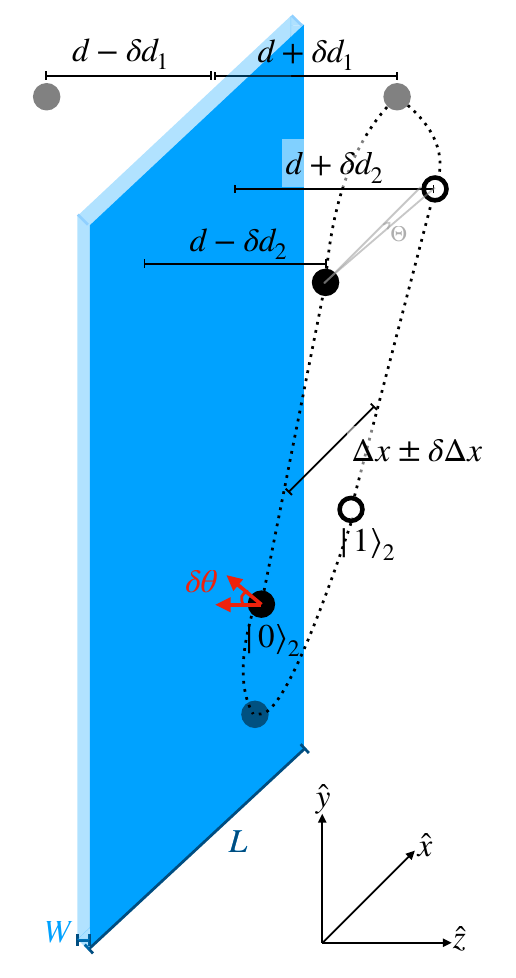}
    \caption{Parallel setup with a conducting plate of size $L$ and width $W$ as illustrated in figure~\ref{fig:setup_plate}.
    The imbalances discussed in section~\ref{sec:phase_fluct} are illustrated. Note that $z(0)=d$ is the initial position, and both $d_1, d_2$ are imbalances in this position due to the distance plate-test mass being different for the test masses, and the distance plate to superposition instances being different between two superposition instances leading to an angle $\theta$, respectively. The imbalance $\delta \Delta x$ is caused by an imbalance in the magnetic field. The red arrow indicates the dipole moment direction, which is assumed to have $\theta=0$ with respect to the direction towards the plate ($\pm \hat{z}$-direction), but which has a fluctuation $\delta \theta$ around this axis.}
    \label{fig:setup_plate_fluc}
\end{figure}

The entanglement is witnessed by performing repeated measurements on the spin states of the spin embedded in the NV-centre of the diamond.
During this process, the test masses need to be prepared repeatedly in the same initial state. 
A small derivation in the initial state results in a change in the final state and therefore yield a different effective entanglement phase.
We consider here imbalances in the distance to the plate (sec.~\ref{subsec:phifluc_d}), the imbalance in the magnetic field gradient (sec.~\ref{subsec:phifluc_dB}) resulting in an imbalance in the superposition size, and an imbalance in the dipole orientation $\theta$ (sec.~\ref{subsec:phifluc_theta}).
These imbalances are illustrated in Fig.~\ref{fig:setup_plate_fluc}.
However, first, we give an estimate of the minimally effective entanglement phase necessary for entanglement to be detectable in the presence of noise from imbalances in the initial conditions, in order to fix our experimental parameters.

\subsection{Minimal Entanglement Phase}
The detection of entanglement is done via a witness which is constructed by repeated measurements on the spin states of the test masses.
Appendix~\ref{app:witness} shows the derivation of the Partial Positive Transpose witness~\cite{Horodecki:2009zz}, which was found to be best for the two-qubit QGEM setup~\cite{Chevalier:2020uvv}.
Its expectation value can be estimated as:
\begin{equation}\label{eq:witness_text}
    \Tr(\mathcal{W}\rho) \approx \gamma t - \frac{1}{2} \Phi_\text{eff} 
\end{equation}
at short time-scales, where $\gamma$ is the decoherence rate, and $\gamma_d$ the dephasing rate. A negative expectation value of the witness is a measure of an entangled state.
In order to find the minimal experimental conditions necessary for detecting entanglement, we consider now what phase $\Delta \phi$ is necessary for the detection of entanglement.

\subsubsection{Shot-noise limited phase}

Fluctuations in the initial conditions of the experiment cause a run-to-run difference in the measurement of the entanglement phase. 
We give a rough estimate of the minimally detectable entanglement phase given a fixed number of measurements.
Assuming that the variation in the initial conditions follows a Normal distribution, we use the shot-noise limited phase uncertainty: 
\begin{equation}
    \Delta\phi_{SN} = \sqrt{\frac{1}{N}} \, ,
\end{equation}
where $N$ is the number of trials. 
We consider the effective entanglement phase to be detectable if it is more than five times ($5\sigma$-rule) larger than the phase uncertainty
\begin{equation}
    \Phi_\text{eff}\geq 5 \Delta\phi_{SN} \, .
    \label{eq:minphase}
\end{equation}
We choose an experimentally realistic number of trials of $N = 10000$~\footnote{
Note that this number of measurements is an estimate based on the shot-noise of the actual number of measurements that one needs to witness entanglement, which is witness-dependent.
}. This corresponds to roughly two weeks of measurement if the time span of one trial is one minute. This is realistic for both the Einstein-elevator type of free fall experiments as well as the particle-launching type of experiments. 
Therefore, the assumed phase uncertainty is $\Delta\phi_{SN} = 0.01$~rad and the minimally detectable phase difference $\Phi_{det} \geq 0.05$~rad.
We note, that in principle, free fall experiments can be repeated almost as fast as the falling time is (a few seconds) and often fully automatically hundreds of thousands of times, which would significantly improve the success chances of the proposed experiments. 

For sources of technical noise relevant to anticipated experiments like magnetic field gradient fluctuations, the initial position fluctuation (see sec. \ref{subsec:phifluc_d}-\ref{subsec:phifluc_dB}), we assume the different quantities do not correlate and vary with a normal distribution $\mathcal{N}(\mu, \sigma)$ around their nominal position $\mu$ with a (conservatively) chosen experimentally based standard deviation $\sigma$. The final effect of these fluctuations on the determined phase difference is reduced by the same factor of $1/\sqrt{N}$

We are fully aware of the fact that the desired effect of gravity-induced entanglement does not have to be larger or even of the same order as ``parasitic'' effects as long as those effects can be corrected for by measuring their different dependence on, for instance, the distance between the micro-spheres. Low statistical uncertainty allows to ``trade statistics for systematics'' and detect even minute effects. Still, we stick to our conservative approach and consider the effect of interest to be differentiable from other effects if its phase exceeds the other effects by the minimal detectable phase $\Phi_\text{eff}$ in eq. \eqref{eq:minphase}.

\subsubsection{Decoherence and dephasing}
As can be seen from eq.~\eqref{eq:witness_text}, when witnessing the entanglement, the entanglement phase is counter-acted by the decoherence rate.
As was studied intensively in~\cite{Tilly:2021qef,Schut:2021svd}
the decoherence rate also increases the number of measurements.
Similarly, dephasing effects and environmental noise sources will make the entanglement harder to measure, and increase the necessary number of measurements. 

From the scattering with air molecules and blackbody photons we would expect a decoherence rate of at least $0.05\,\si{\hertz}$~\cite{Chevalier:2020uvv,Nguyen:2019huk,Schut:2021svd,Tilly:2021qef}.
Therefore $\gamma \tau \sim 0.025$ (since most decoherence happens for large $\Delta x$ only the interaction time $\tau$ is taken). 
In order to get an estimate of the required entanglement phase also including decoherence we require that the witness value with decoherence is still $-0.05$, as it was without decoherence based on the shot-noise limited phase. 
Thus we would require an effective entanglement phase $\Phi_\text{eff} \sim 0.10$ rad.
This would correspond to a superposition width of $\Delta x\approx29\,\si{\micro\metre}$ for an initial distance $d = 41\,\si{\micro\metre}$.
Since the shot noise is taken to be $0.01$ rad, and we consider imbalances in the $4$ initial condition, we consider the fluctuation in the phase due to an imbalance to be at most $12.5\%$, which we use to determine the necessary precision of our initial conditions.

From eq.~\eqref{eq:psit_dephasing} and appendix~\ref{app:witness} we see that the dephasing decreases the effective phase and increases the expectation value of the  witness, respectively.
This will also lead to an increase in the number of measurements necessary for detecting entanglement.
The effect of the dephasing is discussed in more detail in sections~\ref{subsec:fluc_tilt} and~\ref{sec:dephasing}. 

In the remainder of this paper, we consider the test mass to be a diamond NV center with $m = 10^{-14} \,\si{\kilogram}$ and $p = 10^{-2} e \, \si{\centi\metre}$, a plate with $W = 1 \, \si{\micro\metre}$, an experimental time of $1\,\si{\s}$, a superposition width of $\Delta x\approx29\,\si{\micro\metre}$ and an initial separation between the particles of $2d + W = 83\,\si{\micro\metre}$.


\subsection{Fluctuation in the entanglement phase due to an imbalance in $d$}\label{subsec:phifluc_d}

\begin{figure}[t]
    \centering
    \includegraphics[width=\linewidth]{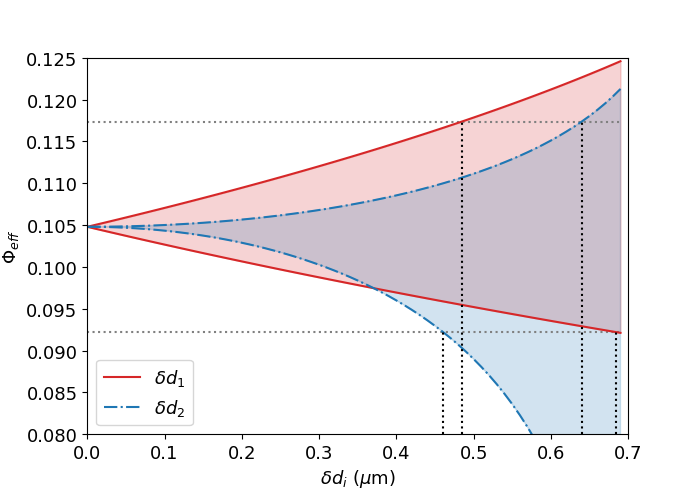}
    \caption{The effective phase accumulated during $1\,\si{\s}$ of experimental time as a function of the variation in the distance $\delta d_i$ ($i=1,2$) in micrometers. 
    The red-shaded region bounded by the solid red lines is the range in effective phase values corresponding to the maximal fluctuation $\delta d_1$ in the distance between the two test masses with respect to the plate. 
    The blue-shaded region bounded by the dash-dotted blue lines is the range in effective phase values corresponding to the maximal fluctuation $\delta d_2$ in the distance between two superposition instances of a single test mass with respect to the plate.
    The horizontal gray dotted lines indicate the $10\%$ deviation from the $\delta d_i = 0$ value, and the vertical black dotted lines show the corresponding values of $\delta d_i$.
    With $m=10^{-14}\,\si{\kg}$ and $\Delta x = 29\,\si{\micro\metre}$.}
    \label{fig:phifluc_d}
\end{figure}

If the test masses are not placed every time exactly at a distance $d$ from the plate, but within a range $d\pm \delta d_1$, with $d_1$ a small fluctuation, then the phase will fluctuate $\Phi_\text{eff} + \delta \Phi_\text{eff}$ run-to-run.
For the starting condition $z(0)=d\pm \delta d_1$, the phase and its fluctuation are plotted as a function of $d_1$ in figure~\ref{fig:phifluc_d} in red.
From the plot, we see that in order to have a fluctuation $\delta \Phi_\text{eff}$ at most $12\%$ of the original phase, for an initial distance of $z(0) = 41 \pm \delta d_1 \, \si{\micro\metre}$, the uncertainty in $d$ is restricted to be $ d_1 < 0.48 \, \si{\mu\metre}$.~\footnote{
The point of closest approximation is also dependent on $d$ and found to be $20 \, \si{\mu\metre}$ for $z(0) = 41.48 \, \si{\mu\metre}$, $16\, \si{\mu\metre}$ for $z(0) = 41 \, \si{\mu\metre}$ and $7\, \si{\mu\metre}$ for $z(0) = 40.52 \, \si{\mu\metre}$.
}
\begin{equation}
    \frac{\abs{\delta \Phi_\text{eff}}}{\Phi_\text{eff}} \leq 0.12\,\, \Rightarrow \,\, \frac{\delta d}{d} \approx 0.01 \, , \qq{} \text{for }d=41\si{\micro\metre} .
\end{equation}
If $d>41\,\si{\micro\metre}$ then larger fluctuations are allowed.
For example if $d=51\,\si{\micro\metre}$, the allowed fluctuation such that $\abs{\delta \Phi_\text{eff}}/\Phi_\text{eff} < 0.12$, is $\delta d_1<1.54\,\si{\micro\metre}$.
However, due to the larger distances, the phase is more than two times smaller.
When placing the masses as close as possible, the room for error in initial conditions is smaller because the distances are smaller and the acceleration towards the plate goes with $1/z^4$ or $1/z^5$, but the accumulated phase is larger exactly because of this reason.
This dependence of the phase on the distance is also the reason for the asymmetry of the red lines in figure~\ref{fig:phifluc_d} around $\Phi_\text{eff}=0.105$.

\subsection{Fluctuation in the entanglement phase due to tilted superposition}\label{subsec:fluc_tilt}
Similarly, if the creation of the spatial superposition is not perfectly parallel with respect to the plate but slightly tilted~\footnote{
The superpositions can be tilted in two ways, in a symmetric way that keeps them parallel ($\Theta_1 = \Theta_2$), and in an asymmetric way such that they are no longer parallel ($\Theta_1 = - \Theta_2$). Both types of tilt were considered here and it was found that the asymmetric tilt provides the upper blue bound line in figure~\ref{fig:phifluc_d}, while the symmetric tilt provides the lower bound blue line.
The reason for this difference can be found by looking at the effective phase (eq.~\eqref{eq:phase}). 
Using a simple geometry argument the asymmetric tilt approximately reduces the second term in the expression for the phase (increasing the effective phase), while the symmetric tilt reduces the first term (reducing the effective phase).
}, this results in a different effective entanglement phase run-to-run.
A tilt resulting in the superposition instances being within a range $d\pm\delta d_2$ away from the plate is considered, the tilt could also be expressed in terms of an angle $\theta$.
For the starting condition $z(0)= 41 \,\si{\micro\metre} \pm \delta d_2$, the phase and its fluctuation are plotted as a function of $\delta d_2$ in figure~\ref{fig:phifluc_d} in blue.
Again, we consider the a maximal phase fluctuation of $12\%$, the figure shows he needed precision is $\delta d_2 < 0.46\,\si{\micro\metre}$, which slightly tightens the precision required previously $\delta d_1< 0.48\,\si{\micro\metre}$~\footnote{
Since there is a difference in attraction towards the plate between the superposition instances, there will be a small enlargement of $\Delta x$. This is not taken into account here since the result $\delta d_2 < 0.49\,\si{\micro\metre}$ suggests the allowed difference in position is so small that this effect is negligible.
}.

The asymmetry of the shaded regions in figure~\ref{fig:phifluc_d} can be explained as follows. 
The shaded regions are asymmetric around the line $\delta d_i = 0$ ($\Phi_\text{eff}=0.105$) because a larger $d$ has a smaller initial attraction force towards the plate and has a larger separation at the end of the experimental time, $t=2\tau_a+\tau$.
This $1/z^4$ or $1/z^5$ dependence of the force  results in an asymmetry in the plot.


The difference in the variations $\delta d_1$ and $\delta d_2$ can be explained from the expression of the effective phase. 
While $\delta d_1$ influences both terms in the effective phase (eq.~\eqref{eq:phase}) approximately in the same way, $\delta d_2$ influences only the second or first term in eq.~\eqref{eq:phase}.
Therefore the variation $\delta d_1$ has less influence on the total effective phase (which is the difference between the first and second terms) compared to $\delta d_2$.

\begin{figure}[h]
    \centering
    \includegraphics[width=1.1\linewidth]{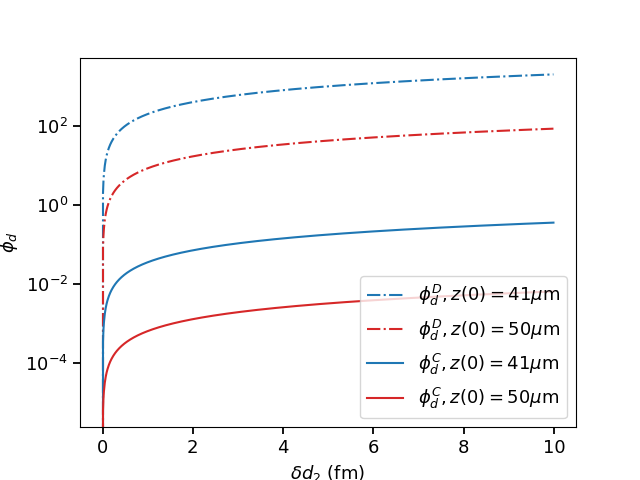}
    \caption{The dephasing accumulated during $1\,\si{\s}$ of experimental time as a function of the variation in the distance $\delta d_2$ in micrometers. 
    With $m=10^{-14}\,\si{\kg}$ and $\Delta x = 29\,\si{\micro\metre}$.
    The solid lines indicate dephasing due to the Casimir interaction, the dash dotted lines indicate dephasing from the dipole-dipole interaction. The blue lines are for an initial separation to the plate of $41$ micron, and the red lines are for $50$ micron.
    }
    \label{fig:dephasing_d2}
\end{figure}

There is an additional constraint on $\delta d_2$ due to the dephasing of the superposition which arises due to the two superposition states having different Casimir-and dipole interactions with the plate and therefore picking up a relative phase.
The interaction of a single superposition with the deflected plate can imprint which-path information on the plate and dephase the test masses.
This source of dephasing is not relevant to the other imbalances because the other imbalances preserve the equidistance of the superposition instances to the plate, which only results in a global phase.
If there is some $\delta d_2>0$ there is a non-global phase as well, which we denote as the dephasing phase $\phi_d$:
\begin{align}\label{eq:dephasing00}
\ket{\psi(\tau)}
    & \propto \frac{1}{\sqrt{2}} \bigg[ e^{i \phi_\text{d}} \ket{0} + \ket{1} \bigg] \, ,
\end{align}
The phase $\phi_d$ is given by the Casimir-and dipole interaction with the plate, $\phi_d\sim (V_C+V_D)t/\hbar$ (see Eqs.~\eqref{eq:casimir_plate_pot},\eqref{eq:dipole_pot}).
The dephasing due to interaction with the point of the plate at the closest approach is:~\footnote{
The expression of the non-global phase is dependent on which way the superposition is tilted. Here we assume that the superposition is tilted such that the $\ket{0}$ state is closest to the plate. Although the expressions~\eqref{eq:dephasing00},~\eqref{eq:dephasing01} would be different if the state $\ket{1}$ were closest to the plate, the total effect would be the same due to the symmetry of the setup.
}.
\begin{align}
    \phi_d &= (\gamma_d^C + \gamma_d^D)t \label{eq:dephasing01} \\
    \gamma_d^C &= \frac{3 c R^3}{8\pi}\left(\frac{\varepsilon-1}{\varepsilon+2}\right) \left[\frac{1}{z_0^4} - \frac{1}{z_1^4} \right] \\
    \gamma_d^D &= \frac{p^2}{16\pi\varepsilon_0\hbar} \left[ \frac{1}{z_0^3} - \frac{1}{z_1^3} \right]
\end{align}
with $\gamma_d^{C,D}$ the dephasing rate due to the Casimir ($C$), dipole ($D$) interaction. Note that this rate is time-dependent due to the time dependence in $z$.
And with $z_0$ and $z_1$ the distance to the plate for the $\ket{0}$ and $\ket{1}$ state, respectively.
Similarly to eq.~\eqref{eq:acc_phase} the total dephasing is found numerically and plotted as a function of $\delta d_2$ in figure~\ref{fig:dephasing_d2}.

In eq.~\eqref{eq:witness_no_approx} we give the witness in terms of the dephasing rate $\phi_d$.
From this expression, we find that for an effective phase of $\Phi_\text{eff} = 0.1$ rad and a decoherence rate of $\gamma = 0.05\,\si{\hertz}$ for a time $\tau$, a dephasing of $\phi_d = 0.2$ rad would increase the witness by approximately a factor two.
In~\cite{Schut:2021svd} this order of increase in the expectation value of the witness led to about an increase in the number of measurements of at least a factor of two.
For a dephasing of $\phi_d = 0.1$ rad the expectation value of the witness stays approximately the same. 

From figure~\ref{fig:dephasing_d2} we see that if we require a maximum dephasing of $\phi_d = 0.1$, the restriction put on the fluctuation $\delta d_2$ is 
$\delta d_2 < 2.8\,\si{\femto\metre}$ from the Casimir interaction and 
$\delta d_2 < 10^{-2}\,\si{\femto\metre}$ from the dipole interaction.
The dominant dipole interaction causes a lot of dephasing, and if the conducting plate setup is to be realized the dipole dephasing has to be mitigated.
Also, the Casimir interaction with the plate which cannot be mitigated puts strict restraints on the precision of the initial conditions, of the order of femtometers.
One can relax these constraints by for example increasing the number of measurements. If we allow $\phi_d = 0.2$, which will increase the number of measurements noticeably, we would require $\delta d_2 < 5.6 \, \si{\femto\metre}$.
Alternatively, we could also increase the mass of the test masses, since $\Phi_\text{eff}$ scales with $m^2$, while $\phi_d$ scales with $m$, this would increase the effective  entanglement phase more relative to the dephasing.
Increasing the separation to the plate also decreases the dephasing, as shown in figure~\ref{fig:dephasing_d2}, however, this also decreases the effective entangling phase.

\subsection{Entanglement phase fluctuation due to Imbalance in $\partial B$}\label{subsec:phifluc_dB}
\begin{figure}[t]
    \centering
    \includegraphics[width=\linewidth]{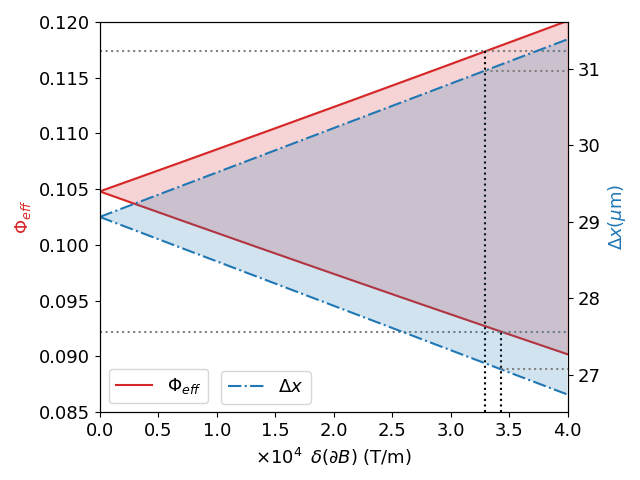}
    \caption{The solid red line shows the effective entanglement phase accumulated during $1\,\si{\s}$ of experimental time as a function of the variation in the magnetic field gradient ($\delta (\partial B)$).
    The (red) left vertical axis shows the effective phase values, with the dotted lines indicating a $12\%$ difference from the $\delta (\partial B) = 0$ value.
    The red-shaded region indicates the phase corresponding to maximal fluctuations given on the horizontal axis.
    The blue dash-dotted lines give the superposition size shown on the (blue) right vertical axis that corresponds to the magnetic gradient on the x-axis. 
    The dotted line indicates the value of $\Delta x$ that corresponds to the $12\%$ phase fluctuations.
    With $m=10^{-14}\,\si{\kg}$, $\partial B = 5\times10^{6}\,\si{\tesla\per\metre}$ (which is $\Delta x = 30\,\si{\micro\metre}$).}
    \label{fig:phifluc_dB}
\end{figure}

Another initial condition that can experience fluctuations run-to-run is the strength of the magnetic field gradient used to create the superposition, $\partial B = \partial B + \delta (\partial B)$. Any foreseeable systematic fluctuation in the magnetic field gradient can influence the phase in each run of the experiment.
For the starting condition $\partial B = 5\times10^5 \pm \delta (\partial B)$, the phase and its fluctuation are plotted in figure~\ref{fig:phifluc_dB}.

From this plot we can see that fluctuations of approximately $\delta \partial (\partial B) = 3.43\times10^4 \, \si{\tesla\per\metre}$, giving the ratio:
\begin{align}
    \frac{\delta (\partial B)}{\partial B} \sim 0.07 \, , \qq{} \text{for }B=5\times10^5\,\si{\tesla\per\metre} .
\end{align}
Similarly to the fluctuation in $d$, a larger $\partial B$ allows larger fluctuations.
Figure~\ref{fig:phifluc_dB} is symmetric around the zero-fluctuation-value since there is a linear dependence of $\Delta x$ on $\partial B$ (see Eq.~\eqref{eq:delta_x}), and for small fluctuations in the superposition size the effective phase $\Phi_\text{eff}$ can  be approximated as:
\begin{equation}
    \Delta \phi = \frac{Gm^2}{\sqrt{d^2+(\Delta x)^2}} \frac{\tau}{\hbar} \left[ 1 + \frac{\delta (\Delta x)}{\Delta x}\right] - \phi \, .
\end{equation}
The dependence on $\Delta x$ and thus $\partial B$ is approximately linear for small variations.

Using Eq.~\eqref{eq:delta_x}, we find the allowed fluctuation in $\Delta x$.
Reading the right vertical axis in figure~\ref{fig:phifluc_dB}, we find that a $12\%$ phase $\Phi_{eff}$, variation corresponds to $\Delta x = 29 \pm 2.0 \,\si{\micro\metre}$.
If the magnetic gradient is caused by the presence of a wire, then its fluctuation is thus caused by a fluctuation in the current of that wire, see~\cite{article,Marshman:2018upe, Zhou:2022jug}.
The fluctuation in the superposition size $\delta (\Delta x)$ is therefore related to the fluctuation in the current $I$~\cite{Marshman:2018upe}
\begin{equation}\label{eq:fluc_deltax}
    0.07 = \frac{\delta (\Delta x)}{\Delta x} \sim \frac{\delta I}{I} \, .
\end{equation} 
Typically, a current density of $J=10^{13}\,\si{\ampere\per\metre\squared}$, which is the ampacity for the copper nanotubes mentioned in footnote~\ref{footnote:current}, matches a current of $I = 7\,\si{\ampere}$~\cite{Subramaniam2013:one}~\footnote{
In~\cite{Subramaniam2013:one} the area of the test material in which the ampacity was measured was $7.2 \times 10^{-13}\,\si{\metre\squared}$, this would correspond to a wire with radius $478\,\si{\nano\metre}$.
}. 
Therefore, given the allowed fluctuation in the superposition size in eq.~\eqref{eq:fluc_deltax}, we can estimate that the fluctuation in the current can at most be $\delta I = 0.48\,\si{\ampere}$. Furthermore, we can also study the fluctuations in the current due to thermal effects in the wire (the Johnson-Nyquist noise) are approximated to be $\delta I \sim 10^{-12}\,\si{\ampere}$ at a room temperature, see~\cite{Marshman:2018upe}. Hence, Johnson-Nyquist noise is well within our estimate on the current fluctuations that satisfy $\abs{\delta\Phi_\text{eff}}/\Phi_\text{eff} \leq 0.12$.

It should be noted that the effective phase is dependent on the protocol for creating the superposition. 
In this paper, we use a very general protocol Eq.~\eqref{eq:delta_x}. 
There are more complicated protocols that can reach the same superposition size using magnetic fields that are experimentally easier to realize, see~\cite{Zhou:2022epb,Zhou:2022frl,Zhou:2022jug,Marshman:2023nkh}.

\subsection{Entanglement phase fluctuation due to imbalance in the dipole moment}\label{subsec:phifluc_theta}

\begin{figure}[t]
    \centering
    \includegraphics[width=\linewidth]{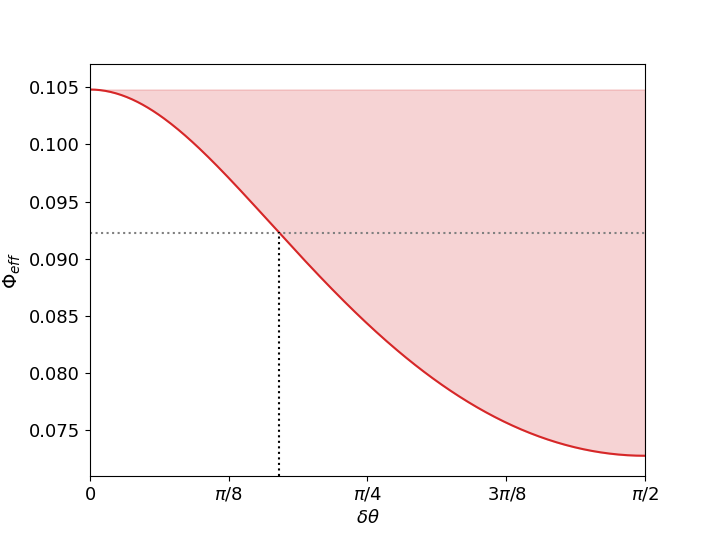}
    \caption{The effective phase accumulated during $1\,\si{\s}$ of experimental time as a function of the variation in the orientation of the dipole moment ($\delta \theta$).
    The red-shaded region bounded by the solid red lines is the range in effective phase values corresponding to the maximal fluctuation $\delta \theta$ on the horizontal axis.
    The horizontal dotted line indicates a $12\%$ difference from the phase at $\delta \theta = 0$, and the corresponding $\delta \theta$-value is indicated by the vertical dotted line. 
    With $m=10^{-14}\,\si{\kg}$, $\Delta x=29\,\si{\micro\metre}$ and $d=41\,\si{\micro\metre}$.}
    \label{fig:phifluc_theta}
\end{figure}

Another initial condition that influences the trajectory of the diamonds and therefore the accumulated effective phase is the orientation of the dipole moments.
So far we have taken $\theta=0$ in Eq.~\eqref{eq:dipole_force}, which gives the `worst-case scenario' in the sense that attraction due to the dipole moment is maximal (and therefore the separation is maximal).
Note that the angle $\theta$ is with respect to the vector going from the test mass towards the plate and is thus aligned with the $-\hat{z}$-direction for the test mass labeled $2$ and the $+\hat{z}$-direction for the test mass labeled $1$ (see Fig.~\ref{fig:setup_plate_fluc}).
We introduce small fluctuations around $\theta=0$ by imagining that the dipoles of the test masses have the same angle $\pm \delta \theta$.
We assume for simplicity that the fluctuation on both test masses is the same. Fig.~\ref{fig:phifluc_theta} shows the range of fluctuations in the effective phase for $\delta \theta \in [0,\pi/2]$.

Since the exerted dipole force goes with $\cos^2(\delta\theta)$, the change in phase is the same for $\pm\delta\theta$.
For $\theta>0$ (but smaller than $\pi/2$) the dipole force decreases and for the starting position $d=41\,\si{\micro\metre}$, the final distance to the plate increases, hence the accumulated entanglement phase decreases.
This can also be seen from figure~\ref{fig:phifluc_theta}. 
The decrease again is non-linear because the dipole force goes with $1/z^4$.
The bound on the dipole moment orientation is given by:
\begin{equation}
    \frac{\abs{\delta \Phi_\text{eff}}}{\Phi_\text{eff}} \leq 0.12\,\, \Rightarrow \,\,
    \delta \theta \approx 0.17 \pi \, ,
\end{equation}
which can be read off from figure~\ref{fig:phifluc_theta} ($0.17\pi \approx \pi/6$).

We have not considered fluctuations in the magnitude of the electric moment, $p$, because we have assumed that the magnitude of the electric moment is due to the internal dipole $p=p_i$, which is constant if the same test mass is used over repeated runs. But a similar analysis can be performed by varying $p$.

\section{Coherence loss due to the conducting plate}\label{sec:dephasing}
In section~\ref{subsec:fluc_tilt} the dephasing due to the plate was found. 
Due to an imbalance in the initial conditions between the two test masses, there can be a net force acting on the plate. 
We find the deflection due to fluctuations $\delta d_1$ and $\delta \theta$. 
Additionally, we estimate dephasing due to thermal fluctuations in the plate.

\subsection{Deflection of the conducting plate}\label{subsec:dephasing_d}

A small uncertainty in the initial placement of the test masses relative to the plate, or in their dipole orientation, will lead to a net force acting on the plate that originates from the difference in the distance-dependent Casimir-and dipole interactions.
The net force causes a deflection in the plate, which is clamped at both ends in the 
$x$ direction. 
We analyze the additional dephasing with respect to sec.~\ref{subsec:fluc_tilt} due to the imbalances discussed previously: $\delta d_1$, and $\delta \theta$.
Note that $\delta (\partial B)$ is independent of the $z(t)$ and does not cause any plate deflection. 

In the linear setup in Fig.~\ref{fig:setup_lin} the net force could be considered a point force acting on the middle of the plate due to the alignment of the superposition with respect to the plate~\cite{vandeKamp:2020rqh}. 
In the parallel setup, the superposition is aligned in a direction parallel to the plate. 
However, we also use the point-source approximation because we consider the length of the sides of the plate (of the order $L\geq1\si{\milli\metre}$) to exceed the superposition width of the test masses (of the order $\Delta x\sim 30\,\si{\micro\metre}$) such that $\Delta x/L\sim 0.03$.
Compared to the size of the plate we approximate the superposition to be point-like.
Additionally, $z(t)>\Delta x(t)$ for any $t$, so although the deflection of the plate is actually in a superposition, we approximate it with a point-source approximation deflection.
Recall that the test masses and plate are in free fall, furthermore, we assume that the test masses are setup up around the mid-point of the plate.
The maximal deflection $\delta z$ at a distance $a$ away from the mid-point of the plate, due to the maximal net point force $F$ at the center of the plate is:~\cite{beams}
\begin{align}
    \delta z_\text{max}
    &= \frac{F_\text{max} (L-2a)^2}{192\,E\,I} (L+4a) \\
    &= \frac{F_\text{max} (L-2a)^2}{16 W^3 E} (1+\frac{4a}{L})\, , \label{eq:deflection}
\end{align}
with $E$ is Young's modulus of elasticity of the plate, $L$ the length of the plate, and $W$ the thickness of the plate.
Eq.~\eqref{eq:deflection} holds for any conducting plate, we specifically consider a Silicon-Nitride plate that is coated with a very thin layer of gold and has a thickness of $1\,\si{\micro\metre}$, and sides of length $L=1\,\si{\milli\metre}$.
We assume that due to the gold layer being very thin, we can take the material properties of the plate to be those of Silicon-Nitride.
Silicon-Nitride has a Young's modulus of $270\,\si{\giga\pascal}$ and a density of $3.1\,\si{\gram\per\centi\metre\cubed}$~\cite{khan2004young}.
The area moment of inertia in the plane of the plate is substituted to get the final result in eq.~\eqref{eq:deflection},
\begin{equation}
    I = \int_{-L/2}^{L/2} \int_{-W/2}^{W/2} z^2 \dd{z} \dd{x} = \frac{1}{12} W^3 L \, .
\end{equation}

The force $F_\text{max}$ is given at the point of closest approach to the center, at the time $\tau+2\tau_a$.
Note that at this time the superpositions have been recombined and the point-force approximation holds true, however, there is no Casimir/dipole dephasing since $\Delta x = 0$.
If there is some maximal imbalance $\pm \delta d_1$ in the initial placement of the test masses, then the maximum force at a distance $a$ from the center is:
\begin{align}
    F_\text{max} &= \,F_\text{C}(z_\text{min}(d+\delta d_1)) -  F_\text{C}(z_\text{min}(d-\delta d_1)) \nonumber \\ &\,\,+  F_\text{D}(z_\text{min}(d+\delta d_1)) -  F_\text{D}(z_\text{min}(d-\delta d_1)) \, ,
\end{align}
with $F_C$ and $F_D$ given in Eqs.~\eqref{eq:casimir_plate} and \eqref{eq:dipole_force}, respectively.

\begin{figure}[t]
    \centering
    \includegraphics[width=\linewidth]{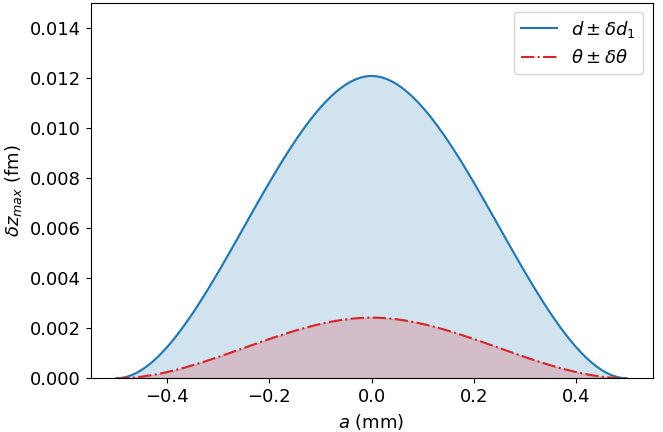}
    \caption{The deflection $\delta z_\text{max}$ in femtometers due to the maximal point force at the time of closest approach $\tau+2\tau_a = 1\,\si{\s}$. For $m = 10^{-14}\,\si{\kg}$, $W=1\,\si{\micro\metre}$, $L=1\,\si{\milli\metre}$, and $E = 270\,\si{\giga\pascal}$ (values for Silicon-Nitride).
    The imbalance in the initial distance to the gold-coated Silicon-Nitride plate is given by $d = 41 \pm 0.48 \, \si{\micro\metre}$ and plotted with the blue solid line. The imbalance in the dipole moment angle is given by $\theta = 0, 0.17\pi$ and plotted with the dash-dotted red line.
    The distance from the midpoint of the plate, $a$, is given on the horizontal axis in millimeters.
    The shaded region is the deflection due to an imbalance smaller than the maximum imbalance.}
    \label{fig:deflection}
\end{figure}

The blue line in figure~\ref{fig:deflection} shows the magnitude of the deflection due to a maximal uncertainty $\delta d_1 = 0.48\,\si{\micro\metre}$ in the initial placement of the test masses, as a function of the distance from the mid-point of the plate, $a$.
The maximal deflection is at the midpoint, $\delta z_\text{max}(a=0) = 0.012 \, \si{\femto\metre}$ (note that the $x$-axis is in millimeters while the $y$-axis is in femtometers, which is $10^{-15}\,\si{\metre}$).

Similarly, a net force can act on the plate due to a difference between the dipole moments.
The net dipole force at $\tau+2\tau_a$ due to a maximal difference $\delta \theta = 0.17\pi$ (meaning $\theta_1 = 0, \theta_2 = 0.17\pi$) causes a deflection of the plate. 
The magnitude of this deflection is plotted in figure~\ref{fig:deflection} in red, as a function of the distance $a$ away from the mid-point of the plate. 
The maximal deflection is $0.002\,\si{\femto\metre}$.
Note that the imbalance $\delta d_2$ at the time $\tau+2\tau_a$ does not cause any deflection.\newline

Using eq.~\eqref{eq:dephasing01} with the new distances due to the deflection of the plate, we find the additional dephasing (compared to Fig. 11). For $\delta d_2 = 2.8\,\si{\femto\metre}$ the additional contribution from the plate fluctuation is given:
\begin{align}
    \Delta\phi_d^C \approx 0.0001 \, ,
\end{align}
which is negligible compared to the phase found in figure~\ref{fig:dephasing_d2} even though this was an overestimation by taking the maximal deflection constant over time.

\subsection{Dephasing due to thermal fluctuations in the plate}\label{subsec:dephasing_th}
Thermoelastic noise, often referred to as just the thermal noise of a membrane, is caused by inevitable local temperature fluctuations around the equilibrium. Temperature variations across the surface of the membrane cause tension and thus vibration  \cite{braginsky1999thermodynamical,kapasi2023direct}. According to the fluctuation-dissipation theorem, there is a corresponding damping process, which is referred to as thermoelastic damping \cite{zener1937internal}. 
\begin{figure}[ht]
    \centering
    \includegraphics[width=\linewidth]{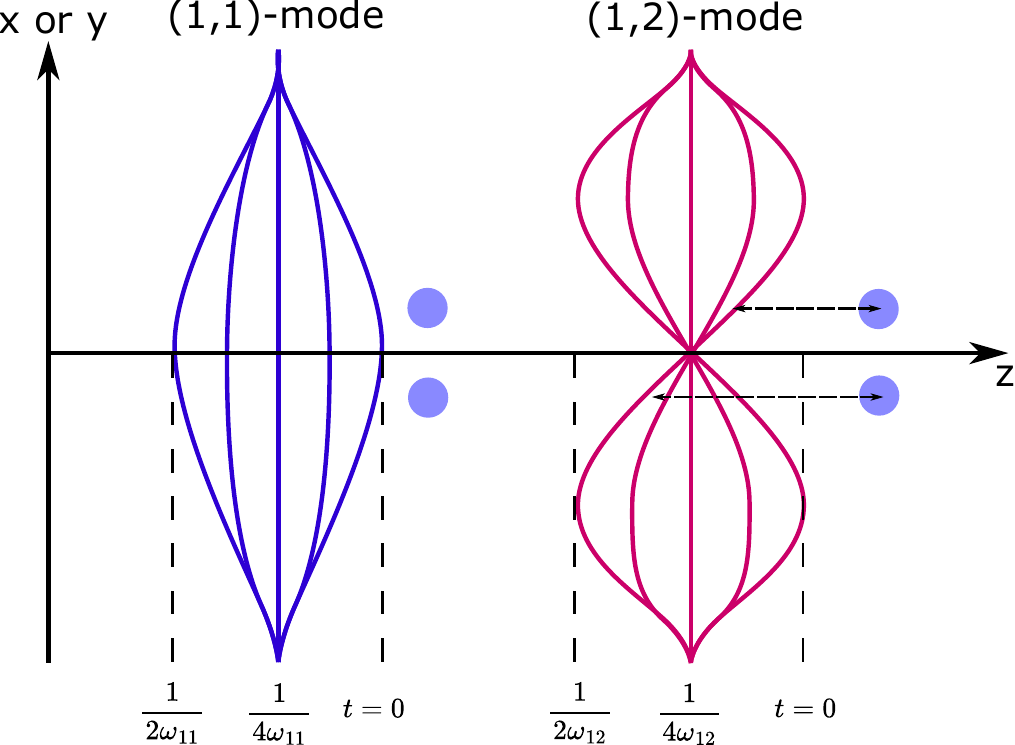}
    \caption{The first two modes of a clamped square plate are drawn schematically together with a spatial superposition state. The spatial superposition state is centered. While the slope of even modes vanishes at the center, odd modes exhibit the necessary asymmetry for a linear effect in $\frac{\Delta x}{L}\ll 1$. The modes are shown several times during one oscillation period $1/\omega_{nm}$ to show that the difference in distance to the membrane changes symmetrically during one whole oscillation period, and thus the effective interaction time is reduced by at least a factor of $2\pi/\omega_{nm}\tau$.}
    \label{fig:membranemodes}
\end{figure}
The thermal motion of the membrane will cause the distance between the spheres and the plate, as well as across the spatial superposition state, to vary. The effect on the trajectory can be neglected as the root-mean-squared (RMS) amplitude is many orders of magnitude smaller than the distance between the spheres and the plate. However, since the phase of the spatial superposition state is much more sensitive to potential variations, it might be affected. We assume that the spheres are centered with respect to the membrane. Since the extent of the spatial superposition ($\Delta x\approx 30 \,\si{\micro\metre}$) is small compared to the size of the membrane ($L=1\,\si{\milli\metre}$), the slopes of even modes at the center will vanish when $\delta d_2=0$~\footnote{
When $\delta d_2 = 2.8\,\si{\femto\metre}$, as found in section~\ref{subsec:fluc_tilt}, the even mode plate deflections (see figure~\ref{fig:membranemodes}) will be maximally of the order $~300~\si{\femto\metre}$. The \textit{additional} dephasing found in the same way as in section~\ref{subsec:dephasing_d} is of the order $\Delta \phi_d^C \sim 2$, which effectively (limiting to effectively one oscillation period as will be discussed later on in this section) 
has a negligible effect of $\Delta \phi^C \sim 10^{-5}$.
}, and their effect will be of at least second order in $\frac{\Delta x}{L}\ll 1$.
The first two modes are illustrated in figure~ \ref{fig:membranemodes}. 
Furthermore, during one complete oscillation period of the membrane, the two components of the spatial superposition state will experience the same overall phase shift, thus reducing the effective interaction time of this dephasing mechanism by at least a factor of $2\pi/\omega_{nm}\tau$ (with $\tau$ the experimental time, and $\omega_{nm}$ the natural frequency of the plate for the mode $(m,n)$). 
Further, in thermal equilibrium, the modes will be exponentially populated, and thus the effect of higher modes decreases. 
We therefore consider only the first odd mode $\omega_{12}$. 

To estimate the root mean squared (RMS) magnitude of the plate deflection, $\delta z_{th}$, due to thermal noise at the temperature $T = 1$K, we use the partition theorem, equate the thermal energy to the potential energy of the plate $k_B T = M\omega_{1,2}^2\delta z_{th}^2$ and solve for $\delta z_{th}$:
\begin{equation}
    \delta z_{th} \leq \sqrt{\frac{k_B T}{M\omega_{nm}^2}} \, .
    \label{eq:deltaz_th}    
\end{equation}
Where $k_B$ is the Boltzmann constant, $T$ is the temperature of the membrane, and $M=\rho WL^2$ is the mass of the membrane. The natural frequencies of a membrane in a vacuum are given by \cite{kinsler1962physical,trushkevych2014laser}:
\begin{equation}
    \omega_{nm} = \frac{1}{2}\sqrt{\frac{\sigma}{\rho}\left(\left(\frac{n}{L}\right)^2+\left(\frac{m}{L}\right)^2\right)} \, ,
\end{equation}
Here $\sigma$ is the biaxial tensile stress, $\rho$ is the density of the material, $L$ is the side length, and $n$, $m$ are the mode numbers. In particular, in the stress-governed regime (membrane), the resonance frequency is independent of thickness. 
The resonance frequencies of a clamped square plate are given  by \cite{reddy2006theory,trushkevych2014laser}:
\begin{equation}\label{eq:omeganm}
    \omega_{nm} = \frac{K_{nm}}{L^2}\sqrt{\frac{EW^2}{12(1-\mu^2)\rho}} \, .
\end{equation}
Here $K_{nm}$ is a mode coefficient ($K_{12}=74.296$, see~\cite{reddy2006theory}), $E = 270\,\si{\giga\pascal}$ the Young's modulus \cite{Norcada} and $\rho=3.1\,\si{\gram\per\centi\metre\cubed}$ the density  of silicon nitride, $W=1\,\si{\micro\metre}$ and $L=1\,\si{\milli\metre}$ the width and side length of the membrane and $\mu=0.2$ the Poisson ratio \cite{khan2004young}.
Finally, we obtain $\omega_{12}=2\pi\times 35.6\,\si{\kilo\hertz}$. 
In \cite{zwickl2008high}, the frequency of a $1\,\si{\milli\metre}\times1\,\si{\milli\metre}$ silicon nitride membrane from Norcada has been measured to be $\omega_{12}=2\pi\times 211613\,\si{\hertz}$. To be more conservative and allow for flexibility in material choice and sample-to-sample difference of membranes, we stick to the previously used model of a clamped square plate in eq~\eqref{eq:omeganm}, 
which gives $\omega_{12}=2\pi\times 35.6\,\si{\kilo\hertz}$.

For the RMS distance fluctuation at $T=1\si{\kelvin}$ eq.~\eqref{eq:deltaz_th} gives $\delta z_{th} = 298\,\si{\femto\metre}$. 
We include the geometric reduction of order $\Delta x /L \approx 0.03$ ($\Delta x = 30\,\si{\micro\metre}$, $L = 1 \,\si{\milli\metre}$). 
The effective differential distance to the surface (analogous to $\delta d_{2}$) is approximately $9\,\si{\femto\metre}$. 
Also including the fact that the interaction time for the asymmetric part of the interaction is limited to effectively one oscillation period $2\pi /\omega_{12}$ of the membrane, the effect on the random phases is further suppressed by a factor of $3.6\times 10^4$, corresponding to only an effective $\delta d_{2}$ of $2.5\cdot 10^{-19}\,\si{\metre}$. Thus the effect can be neglected (see figure~\ref{fig:dephasing_d2}).

We note that the silicon nitride membranes must be coated with a conductor, for example, gold, to effectively act as a Casimir screen, but the coating can be made much thinner than the silicon nitride layer, and we assume the mechanical properties to change only slightly.

\section{Conclusion \& Discussion}\label{sec:conc}

The experimental protocol considered in this paper is that of two free-falling test masses that are separated by a conducting plate and which are free to move towards the plate due to the Casimir-and dipole interaction. This would mean that we have to prepare at least two traps for the experiments, one for releasing the nano-diamond, and the other for capturing.
The conducting plate prohibits quantum electromagnetic interaction between the test masses, due to which the condition for the minimal distances can be removed to some extent.
The minimal distance was first introduced in order to keep the Casimir force subdominant compared to the gravitational force between the test masses. 
However, in free fall there is still some (mass-dependent) initial separation required in order for the test masses to not fall into the plate within one second of experimental time.
The minimal distance for different masses such that the particle (with $0.48\,\si{\micro\metre}$ fluctuation) does not hit the plate, and the superposition size needed to find an entanglement phase of $0.10$ is given in table~\ref{table:freefall_dx} for different masses.
Compared to the absence of the conducting plate, we are able to bring the test masses closer and as a result, allow smaller values of $\Delta x$.

\begin{table}[t]
\bgroup
\def\arraystretch{1.5}
\begin{tabular}{|| c | c | c | c ||} 
 \hline
 \hspace{1mm} Mass (kg) \hspace{1mm} & \hspace{1mm} $2d+W (\si{\micro\metre}$) \hspace{1mm} &  
 \hspace{1mm} $\Phi_\text{eff}$ \hspace{1mm} & \hspace{1mm} $\Delta x\, (\si{\micro\metre})$ \hspace{1mm} \\ [0.5ex] 
 \hline\hline
 $10^{-14}$ & 83 & 0.10 & 29
 \\
 \hline
 $10^{-13}$ & 53 & 0.10 & 1.4
 \\  
 \hline
 $10^{-12}$ & 35 & 0.10 & 0.08
 \\  
 \hline
\end{tabular}
\egroup
\caption{Experimental parameters required to get the desired entanglement phase for two test masses in free fall separated by a superconducting plate. The initial distance ($2d+W$) is taken such that the test masses within $1\,\si{\s}$ experimental time do not collide with the plate and such that a fluctuation of at least $0.5\,\si{\micro\metre}$ is allowed.
For $p = 10^{-2} e \, \si{\centi\metre}$ ($\theta=0$) and a plate with $W = 1 \, \si{\micro\metre}$. The final distance to the plate at the end of the free fall, $z(\tau+2\tau_a)$, are $16, 11, 9 \,\si{\micro\metre}$ for masses $10^{-14}, 10^{-13}, 10^{-12}\,\si{\kilogram}$ for the initial distance given in the table.}
\label{table:freefall_dx}
\end{table}

For a diamond test mass with an embedded NV-center of $m=10^{-14}\,\si{\kilogram}$, instead of a minimal distance between the test masses of $\approx 147\,\si{\micro\metre}$ in the absence of the conducting plate, we now initialize the system with $d = 83\,\si{\micro\metre}$ and after $1\,\si{\s}$ of experimental time during which the test masses are attracted towards the conducting plate due to the Casimir-and dipole force, the smallest separation of the test masses is approximately $33\,\si{\micro\metre}$.
This setup allows for a smaller separation and thus a larger quantum gravitational interaction.
The entanglement phase is enhanced and the experimental parameters can be somewhat relaxed compared to the setup without the plate.
For a mass of $10^{-14}\,\si{\kilogram}$ a superposition size of $29\,\si{\micro\metre}$ is enough for an entanglement phase of $\order{10^{-1}}$.
In the absence of the conducting plate the same order can be reached for a superposition size of $\sim 100\,\si{\micro\metre}$ (see figure~\ref{fig:lin_par_comparison}).

\begin{table}[t]
\bgroup
\def\arraystretch{1.3}
\begin{tabular}{|| c  | c ||} 
 \hline
 \hfil parameter value  & \hfil fluctuation value \\ [0.5ex] 
 \hline\hline
    \hfil $d = 41 \,\si{\micro\metre}$ & \hfil \begin{tabular}{@{}c@{}} $\delta d_1 = 0.48 \,\si{\micro\metre}$ \\ $\delta d_2 = 2.8 \,\si{\femto\metre}$ \end{tabular}  
    \\ 
    \hline
    \hfil $\partial B = 5 \times 10^5\,\si{\tesla\per\metre}$ & \hfil $\delta (\partial B) = 3.4 \times 10^4 \,\si{\tesla\per\metre}$
    \\
    \hline
    \hfil $\Delta x = 29 \, \si{\micro\metre}$ & \hfil $\delta (\Delta x) =  1.6 \si{\micro\metre}$
    \\
    \hline
    \hfil $\theta = 0$ & \hfil $\delta \theta = 0.17\pi$
    \\  
    \hline
\end{tabular}
\egroup
\caption{The imbalance $\delta d_1$ in the distance between the test masses and the plate, the imbalance in the distance between two superposition instances $\delta d_2$ caused by the superpositions being not perfectly parallel to the plate, the imbalance $\delta(\Delta x)$ in the superposition size caused by an imbalance $\delta(\partial B)$ in the magnetic gradient, and an imbalance $\delta\theta$ in the orientation of the dipole moment, are illustrated in figure~\ref{fig:setup_plate_fluc}. 
For a diamond NV centre with $m = 10^{-14} \,\si{\kilogram}$, $p = 10^{-2} e \, \si{\centi\metre}$ and a plate with $W = 1 \, \si{\micro\metre}$.}
\label{table:imbalances}
\end{table}

Since the witnessing of entanglement requires repeating the experiment and performing the measurements many times, we also considered several types of imbalances in the initial conditions of the setup.
In appendix~\ref{app:witness} we found that the entanglement can be witnessed if $\Phi_\text{eff}>2\gamma t$.
For a decoherence rate $\gamma<0.05\,\si{\hertz}$, the results for the maximum deviation in the initial conditions that is allowed in order for the entanglement to be witnessable are summarised in table~\ref{table:imbalances}.
We have also considered the dephasing due to the net force that the imbalances can exert on the plate and found that it was negligible (section~\ref{sec:dephasing}).

In order for the experimental protocol proposed here to be realized one needs to very precisely know the initial positions.
Furthermore, as a worst-case scenario, we have considered here the dipole moment orientation to be towards the plate $\pm$ some fluctuation around this axis.
An opposite dipole moment orientation would result in a repulsive force between the plate and test mass and would thus require an initial positioning of the test mass close to the plate such that the total trajectory is approximately opposite compared to what is illustrated in figure~\ref{fig:trajectory2}.
Therefore most important in this setup and in determining the best initial separation is controlling the dipole moment direction and mitigating the dephasing due to the dipole, which dominates any other phase. 
%

In conclusion, adding the conducting plate in a parallel configuration enhances the entanglement signal, but one needs to control the dipole in order for this setup to yield better results.
In this paper we have estimated the strength of the dipole moments based on previous measurements with silica microspheres. Fabrication and engineering of custom test masses with reduced permanent dipole moments could be a promising approach towards mitigating these background effects. 

\section*{Acknowledgements} \label{sec:acknowledgements}
MS is supported by the Fundamentals of the Universe research program at the University of Groningen. 
A.G. is supported in part by NSF grants PHY-2110524 and PHY-2111544, the Heising- Simons Foundation, the John Templeton Foundation, the W. M. Keck Foundation, and ONR Grant N00014-18- 1-2370. 
S.B. would like to acknowledge EPSRC grants (EP/N031105/1, EP/S000267/1, and EP/X009467/1) and grant ST/W006227/1.
AM’s research is funded by the Netherlands Organisation for Science and Research (NWO) grant number 680-91-119.

\bibliography{casimir.bib} 
\bibliographystyle{ieeetr}

\newpage
\onecolumngrid
\appendix
\section{Witnessing entanglement}\label{app:witness}
To experimentally witness the entanglement, we will construct a witness $\mathcal{W}$. 
From the separability condition of the non-entangled states, one can construct the Positive Partial Transpose (PPT) witness~\cite{PhysRevA.46.4413,Horodecki:2009zz}.
This witness was found to be more optimal for this type of experiment compared to for example a CHSH-type witness~\cite{Chevalier:2020uvv,PhysRevA.72.012321,Schut:2021svd}.

The PPT witness gives a criterion for the separable states, which results in the condition that a state is separable if its partial transpose has no negative eigenvalues (also called the Peres–Horodecki criterion).
Therefore, constructing the PPT witness is given by:
\begin{equation}
    \mathcal{W} = (\ket{\lambda_{-}}\bra{\lambda_{-}})^{T_i} \, , \label{eq:ppt}
\end{equation}
where $\ket{\lambda_{-}}$ is the eigenvector corresponding to the minimal eigenvalue of $\rho^{T_i}$ (the partial transpose of $\rho$), provides a way to test if a state is non-separable. 
Because all separable states satisfy:
\begin{align}
    \Tr(\mathcal{W}\rho) 
    &= \Tr(\ket{\lambda_{-}}\bra{\lambda_{-}}\rho^{T_i}) =  \Tr(\bra{\lambda_{-}}\rho^{T_i}\ket{\lambda_{-}}) = \lambda_{-} \geq 0 \, .
\end{align}
Hence, by using the cyclic property and the invariance under the partial transpose of the trace, we obtain:
\begin{equation}
    \text{If } \Tr(\mathcal{W}\rho) < 0 \text{ then } \rho \text{ is non-separable.} 
\end{equation}
This is a necessary and sufficient condition for the entanglement criterion for the qubit-qubit system considered here in the context of the QGEM experimental protocol.

By including the decoherence rate, $\gamma$, in the same way as was done in~\cite{Schlosshauer:2019ewh,Schlosshauer:2014pgr,Schut:2021svd},  and including the dephasing rate to be  $\gamma_d$ as given in section~\ref{sec:dephasing}, we will have to define the density matrix in terms of  the final spin states, given in eq.~\eqref{eq:psit}. The eq.~\eqref{eq:psit} gives us the density matrix, $\rho = \ket{\Psi(t)}\bra{\Psi(t)}$, for the parallel case without taking into account the decoherence. However, now by taking the effect of decoherence, we get:
\begin{align}
\rho = \frac{1}{4}
    \begin{pmatrix}
        1 & e^{-i\Delta\phi+i\Delta\phi_d - \gamma t} & e^{-i\Delta\phi- \gamma t} & e^{i\Delta\phi_d - 2\gamma t} \\
        e^{i\Delta\phi-i\Delta\phi_d - \gamma t} & 1 & e^{-i\Delta\phi_d - 2 \gamma t} & e^{i\Delta\phi - \gamma t} \\
        e^{i\Delta\phi - \gamma t} & e^{i\Delta\phi_d - 2 \gamma t} & 1 & e^{i\Delta\phi+i\Delta\phi_d - \gamma t} \\
        e^{-i\Delta\phi_d - 2 \gamma t} & e^{-i\Delta\phi
        - \gamma t} & e^{-i\Delta\phi-i\Delta\phi_d - \gamma t} & 1
    \end{pmatrix}
\end{align}
Here $\Delta\phi$ is the entanglement phase given in eq.~\eqref{eq:phase}, $\Delta\phi_d$ is the dephasing described in section~\ref{sec:dephasing} and $\gamma$ is the decoherence rate. 

We present a simplified witness based on the PPT criterion, based on the eigenvectors corresponding to the minimal eigenvalue of the partial transpose of the density matrix.
Ref.~\cite{Chevalier:2020uvv} showed the expansion of the witness given in eq.~\eqref{eq:ppt} for the linear setup. 

Ref.~\cite{Schut:2023hsy} showed the same expansion for the parallel setup.
The decomposition of this witness in a Pauli basis for the parallel setup was found to be:
\begin{equation}
    \mathcal{W} = \left(1 \otimes 1 - X \otimes X + Z \otimes Y + Y \otimes Z \right) \, ,
\end{equation}
where $I, X, Y, Z$ correspond to the identity matrix and the Pauli matrices, respectively.

By using that the trace is a linear map, the expectation value of the witness, $\Tr(\mathcal{W}\rho)$, can then be simplified for the parallel and the linear setups, respectively: (see~\cite{Chevalier:2020uvv} for the linear setup):
\begin{align}
    \Tr(\mathcal{W}\rho) 
    &= \Tr(\rho) \mp 2 \Im(\rho_{12}) \mp 2 \Im(\rho_{13}) - 2 \Re(\rho_{14}) - 2 \Re(\rho_{23}) \pm 2 \Im(\rho_{24}) \pm 2 \Im(\rho_{34})
\end{align}
Using this identity, and performing the expansion for a small-time expansion appropriate for the experimental protocol, gives the result for the parallel setup:
\begin{align}
\Tr(\mathcal{W}\rho)
    &= 1 - e^{-\gamma t} \left( - \sin(\Delta\phi)[ 1+\cos(\Delta\phi_d) ] + \cos(\Delta\phi_d)e^{-\gamma t)}  \right) \label{eq:witness_no_approx} \\
    &\approx 1 - (1-\gamma t) \left[ - 2 \Delta\phi + (1-\gamma t) \right] + \frac{\Delta \phi_d^2}{2} (1-\gamma t)\left[  \Delta \phi - (1-\gamma t) \right]
    \approx \gamma t + \Delta \phi +\order{t^2} \, . \label{eq:witness}
\end{align}
where the approximation is obtained by expanding the expression around the time, $t=0$, and keeping the first-order terms in the expansion.

Note that for the parallel setup, the witness is given by: $\gamma t + \Delta \phi$ but always $\Delta \phi<0$, so it is equivalent to taking $\gamma t - \abs{\Delta \phi}$, which was found in Ref.~\cite{Chevalier:2020uvv}, for the linear setup.

The expectation value of the witness is negative, and therefore, the entanglement can be measured if $\abs{\omega} > \gamma $, where $\omega t = \Delta \phi$.
A small dephasing rate is negligible in this description. 
The condition for the entanglement can be rewritten in terms of the effective phase:
\begin{equation}\label{eq:witness_condition}
    \Phi_\text{eff} > 2 \gamma t
\end{equation}
For simplicity, we may neglect here the sign difference between the parallel and the linear case, and in this work, the absolute value of $|\omega|$ is implied.

\end{document}